\documentclass[]{aa} 
\usepackage{natbib}
\bibpunct{(}{)}{;}{a}{}{,} 
\usepackage[varg]{txfonts}
\usepackage{graphicx}
\usepackage{color}
\usepackage{caption}
\usepackage{subcaption}
\usepackage{wasysym}
\usepackage{amsmath}
\usepackage{morefloats}
\usepackage{balance}
\pdfoutput=1

\begin{document}

\title{Modeling gravitational instabilities in self-gravitating protoplanetary disks with adaptive mesh refinement techniques}
\titlerunning{Modeling GI in self-gravitating protoplanetary disks with AMR techniques} 
\authorrunning{Lichtenberg \& Schleicher}

\author{
Tim Lichtenberg\inst{\ref{inst1},\ref{inst2}, \ref{inst4}}
\and 
Dominik R.\,G. Schleicher\inst{\ref{inst3},\ref{inst4}}
}

\institute{
Institute for Astronomy, ETH Z\"urich, Wolfgang-Pauli-Strasse 27, 8093 Z\"urich, Switzerland
\label{inst1} 
\and
Institute of Geophysics, ETH Z\"urich, Sonneggstrasse 5, 8092 Z\"urich, Switzerland 
\label{inst2} 
\and 
Departamento de Astronom\'ia, Universidad de Concepci\'on, Av. Esteban Iturra s/n Barrio Universitario, Casilla 160-C, Chile
\label{inst3}
\and
Institut f\"ur Astrophysik, Georg-August-Universit\"at G\"ottingen, Friedrich-Hund-Platz 1, 37077 G\"ottingen, Germany  \\
\email{tim.lichtenberg@phys.ethz.ch} \label{inst4}
}

\date{Received July 4, 2014, accepted April 27, 2015}

\abstract
{
The astonishing diversity in the observed planetary population requires theoretical efforts and advances in planet formation theories. The use of numerical approaches provides a method to tackle the weaknesses of current models and is an important tool to close gaps in poorly constrained areas such as the rapid formation of giant planets in highly evolved systems. So far, most numerical approaches make use of Lagrangian-based smoothed-particle hydrodynamics techniques or grid-based 2D axisymmetric simulations.

We present a new global disk setup to model the first stages of giant planet formation via gravitational instabilities (GI) in 3D with the block-structured adaptive mesh refinement (AMR) hydrodynamics code \textsc{enzo}. With this setup, we explore the potential impact of AMR techniques on the fragmentation and clumping due to large-scale instabilities using different AMR configurations. Additionally, we seek to derive general resolution criteria for global simulations of self-gravitating disks of variable extent.

We run a grid of simulations with varying AMR settings, including runs with a static grid for comparison. Additionally, we study the effects of varying the disk radius. Adopting a simple thermodynamical profile, corresponding to a marginally stable disk ($Q_{init}=1$), we validate the numerical robustness of our model for different spatial extensions, from compact to larger, extended disks. The physical settings involve disks with $R_{disk} = 10, 100$ and $300$ AU, with a mass of $M_{\text{disk}} \approx 0.05 M_{\odot}$ and a central object of subsolar mass ($M_{\star} = 0.646 M_{\odot}$). To validate our thermodynamical approach we include a set of simulations with a dynamically stable profile ($Q_{init}=3$) and similar grid parameters.

The development of fragmentation and the buildup of distinct clumps in the disk is strongly dependent on the chosen AMR grid settings. By combining our findings from the resolution and parameter studies we find a general lower limit criterion to be able to resolve GI induced fragmentation features and distinct clumps, which induce turbulence in the disk and seed giant planet formation.
Irrespective of the physical extension of the disk, topologically disconnected clump features are only resolved if the fragmentation-active zone of the disk is resolved with at least $100$ cells. The latter corresponds to a minimum requirement for all global disk setups.

Our simulations illustrate the capabilities of AMR-based modeling techniques for planet formation simulations and underline the importance of balanced refinement settings to reproduce fragmenting structures. The clumps in our models are migrating inward and are eventually destroyed because of tidal disruptions, reflecting the absence of radiative feedback from the central star, which may stabilize the clumps on larger scales. We expect that the inclusion of additional physics, like a radiation transport mechanism and the formation of sink particles, will provide a detailed framework to study the formation of planets via gravitational instabilities in a global disk view.
 
}


\maketitle

\section{Introduction}
Since the dawn of observational exoplanet detection in recent decades the ever-growing number and variety of planetary species
gives rise to renewed theoretical efforts for understanding the formation of these bodies \citep[e.g.,][]{1997Sci...276.1836B,2000Icar..143....2B}. The ultimate goal of achieving a consistent explanation for the planetary population \citep[e.g.,][]{2014prpl.conf..691B} can only be adressed by pushing observational and theoretical tools to their limits. 

The two major theories of planet formation, core accretion (CA) and formation via gravitational instabilities (GI), have the potential to explain a wide variety of planetary species. Whereas it is widely accepted that terrestrial planets are formed via a bottom-up model like CA \citep[e.g.,][]{2006ApJ...644.1223R,2012AREPS..40..251M,2014prpl.conf..595R}, the formation of more massive objects, like gas giants, can in principle be understood in terms of CA \citep{2014prpl.conf..643H} or GI \citep{2003ASPC..294..281M}. Recent efforts \citep[e.g.,][]{2009ApJ...695L..53B} are dedicated to a synthesis of both models in a more general framework of planet formation. Scenarios based on GI, however, can be particularly relevant for the formation of planets around metal-poor stars, where dust grain coagulation becomes increasingly difficult \citep[e.g.,][]{2010Sci...330.1642S,2012EGUGA..1411532J}, and the buildup of supermassive and Population III stars \citep[e.g.,][]{2014MNRAS.445.1549I,2015MNRAS.449...77L,2015arXiv150406296S}.

In this work we aim for an investigation of the capabilities of an alternative technical treatment of GI to achieve a better understanding of the buildup of massive planets. This includes gaseous planets like Jupiter and Saturn in our own solar system, as well as similar or even more massive planets in extrasolar systems \cite[e.g.,][]{2010Sci...327..977B} or planets around exotic configurations like in the pre-cataclysmic binary system NN Serpentis. The latter system is an example for the class of evolved binaries, for which planets were detected \citep{2013A&A...549A..95Z,2013A&A...555A.133B}. NN Serpentis recently underwent detailed investigations \citep[e.g.,][]{2004A&A...428..181H,2011AIPC.1331..281H,2012MNRAS.425..749H,2014MNRAS.437..475M,2014A&A...562A..19V,2014MNRAS.438L..91P}, indicating a second-generation scenario where the planets were formed after a common envelope (CE) event. Subsequently, \cite{2014A&A...563A..61S} argued for these planets to be consistent with the formation via GI and introduced a semianalytical model, which further motivates studies in compact self-gravitating disks. In the NN Serpentis scenario, the age of the white dwarf of $\lessapprox 10^6$ yr provides an upper limit on the timescale for planet formation, therefore favoring GI with respect to CA and indicating a formation scenario in a very compact environment (< $10$ AU). 

So far, the hydrodynamical treatment of GI in protoplanetary disks mostly relies on a description via smoothed-particle hydrodynamics (SPH) \citep[e.g.,][]{2007ApJ...661L..77M} or axisymmetric \citep[e.g.,][]{2003ApJ...599..577B,2007ApJ...665.1254B,2011MNRAS.410..559M} static grid approaches without adaptive mesh refinement (AMR). Only a minority of grid approaches employ a mechanism for adaptive and dynamic refinement of the initial mesh \citep[e.g.,][]{2006A&A...450.1203P}. In the case of Cartesian grids, the only attempt we know about is the project of \cite{2008ASPC..398..243M} and \cite{2010EAS....42..267G}, who compared \textsc{flash} and \textsc{gasoline} simulations of a self-gravitating disk. In general, the use of more than one numerical approach yields the advantage of ruling out systematic errors, introduced by the chosen numerical treatment \citep[e.g.,][]{2007MNRAS.380..963A}. Advantages of grid codes employing finite volume methods in general are their built-in option to conserve mass and momentum. Regarding the treatment of astrophysical phenomena they offer a precise modeling of turbulence in hydrodynamical instabilities \citep{2007MNRAS.380..963A} and a more robust treatment of shocks in comparison with SPH approaches \citep{2010MNRAS.406.1659P}. Additionally, classical SPH approaches tend to overrate viscosity parameters \citep{2008JCoPh.22710040P,2012JCoPh.231..759P}, which is not the case for grid-based implementations. However, recent developments in the SPH community extenuate many of these problems \cite[see, e.g.,][]{2012MNRAS.422.3037R,2013MNRAS.428.2840H}. In all numerical codes the chosen geometry dictates the implementation and discretization of the physical equations. In the case of disk simulations, one could argue that the natural geometry of choice would be axisymmetric. However, we see some principle advantages in the use of Cartesian geometry, which motivates comparative studies. For example such codes do not include a built-in singularity (i.e., the center of origin) and handle all computed regions equally. The latter means there is no built-in refinement in inner regions, which might avoid symmetry-dependent solutions. Including the use of AMR schemes enables us to specificially choose regions of greater interest and neglect others. This can be particularly relevant when fragmentation occurs because of turbulent motion in the outer disk parts and when the deviations from axisymmetry are more prominent.

We use the Eulerian block-structured AMR hybrid code \textsc{enzo} \citep[][\textsc{enzo-project.org}]{2004astro.ph..3044O,2014ApJS..211...19B}, which enables massively parallel simulations and supports a wide variety of astrophysical problems, i.e., hydrodynamics, ideal and non-ideal magnetohydrodynamics, and N-body dynamics, including self-gravity for fluids and gas chemistry. The fluid flow in \textsc{enzo} is evolved using a finite-volume discretization, the ideal gas dynamics calculations are treated with the piecewise parabolic method (PPM), which is a higher-order Godunov scheme \citep{1995CoPhC..89..149B} with an accurate piecewise parabolic interpolation and a non-linear Riemann solver for shock conditions. The solver is third-order accurate in space and second order accurate in time for fixed time stepping and formally second-order accurate for variable time stepping. \textsc{enzo} has been tested on a variety of typical fluid flow test problems \citep{2014ApJS..211...19B}.

The code employs the implementation of a (structured) adaptive mesh refinement (AMR) scheme, first developed by \cite{1989JCoPh..82...64B}, which enables higher resolution than affordable via uniform grids by introducing additional finer mesh structures on a coarse uniform grid when necessary (which can be defined with a variety of conditions). This scheme utilizes an adaptive hierarchy of rectangular grid patches, which cover a snippet of space with a certain resolution, refined from the top to the bottom of the hierarchy. When the solution evolves and interesting regions develop, finer meshes are placed within the coarse grid, enabling a higher resolution of these structures and not draining too much computing power for less interesting regions.
This facilitates a variety of options for modeling GI in protoplanetary disks, in which we need to deal with dynamical effects on a large range of spatial scales.

We perform 3D simulations of gravitational instabilities in protoplanetary disks with varying disk radius and resolution settings using adaptive mesh refinement techniques. In Section \ref{sec:methodology} we describe the physical and numerical setup of the simulations. In Section \ref{sec:resolution} we compare the initial and evolved state of stable and unstable disks and investigate the effect of different refinement settings. In Section \ref{sec:parameter} we verify the setup for various disk radii and demonstrate its ability to model systems of a wide range of disk extensions. Finally, we summarize our findings in Section \ref{sec:conclusions}.

\section{Numerical experiments}
\label{sec:methodology}

\subsection{Numerical setup}
\label{sec:NS}

For a complete description of \textsc{enzo}'s code structure we refer to \citet{2014ApJS..211...19B}. To provide a sufficient explanation of our numerical methods, however, we briefly outline the major numerical recipes used in the calculations.

\subsubsection*{Hydrodynamic equations} 

In our simulations, we use the piecewise parabolic method (PPM), which was originally developed by \citet{1984JCoPh..54..174C} and was adapted for the \textsc{enzo} code by \citet{1995CoPhC..89..149B} in a direct Eulerian fashion. In this implementation, the governing equations are dimensionally split and rewritten (as example in one dimension) in conservative form as
\begin{align}
\frac{\partial \rho }{\partial t} +  \frac{\partial \rho v}{\partial x} & = 0,\\
\frac{\partial \rho v}{\partial t} + \frac{\partial \rho v^2}{\partial x} +  \frac{\partial p}{\partial x} & = \rho g, \\
\frac{\partial \rho E}{\partial t} + \frac{\partial \rho v E}{\partial x} & = \rho v g,
\end{align}
in which $x$ and $v$ refer to the one-dimensional position and velocity of the gas, $g$ to the acceleration at the cell center, $\rho$ to the gas density, and $E$ to the total fluid energy density. The one-dimensional solution to these equations can be found by computing the effective left and right states at each grid boundary. This is done by constructing a piecewise parabolic solution of density, velocity and total energy and then averaging over the distance that each wave can travel in the specific time step. Then a Riemann problem with these effective left and right states is solved and all quantities are updated. The standard Riemann solver for the PPM method in \textsc{enzo} approximates the appearing rarefaction waves, which are traveling left and right, as shocks. The solution is then found using an iterative approach.
For the case of occurring numerical problems due to the Riemann solver we used, \textsc{enzo} employs a fallback mechanism. This allows the code to switch the Riemann solver to the more diffusive HLL solver at particular interfaces, where negative densities or energies occur \citep[see, e.g.,][]{2009ApJ...691.1092L}.
Finally, the fluxes in between the grid patches are constructed. The full, one-dimensional mathematical description and the interpolation methods are given in \cite{1984JCoPh..54..174C} and \citet{1995CoPhC..89..149B}.

\subsubsection*{Dual energy formalism} 

As long as the structures in the simulations are well resolved and the Mach numbers are on a reasonable scale (i.e., $Ma < 100$), the aforementioned solver works well. However, if hypersonic bulk flows with $E_{\text{therm}}/E_{\text{tot}} \sim 10^{-3}$ appear in the simulation, the numerical situation becomes disastrous: the ratio of kinetic energy $E_k$ to gas internal energy $e$ can approach numbers that are much too high. Hence, the pressure (proportional to $E - E_{kin}$) is the difference between two extremely large numbers, which would be a rather disadvantageous numerical situation with major sources for errors, especially for the temperature distribution. To overcome this problem, \textsc{enzo} separately solves the internal energy equation
\begin{align}
\frac{\partial e}{\partial t} + \mathbf{v} \cdot \mathbf{\nabla} e = - \frac{p}{\rho} \mathbf{\nabla} \cdot \mathbf{v},
\label{eq:intE2}
\end{align}
with thermal energy density $e$ and pressure $p$, which is then used in hypersonic flows for calculating the pressure and thus the temperature distribution only. The gas dynamics (i.e., the hydrodynamic routines) should be unaffected to avoid introducing new errors. This is achieved by choosing a selection criterion for the pressure and thus separating both formalisms via
\begin{align}
p = \left\{
\begin{array}{l l}
& p(\gamma - 1)(E - \mathbf{v}^2/2), \; (E - \mathbf{v}^2/2)/E > \eta  \\ 
& p(\gamma - 1)e, \hspace*{1.69cm}  (E - \mathbf{v}^2/2)/E < \eta
\end{array}
\right\}.
\end{align}
If the barrier parameter $\eta$ is chosen to be small enough, the dynamical effect of the dual energy formalism is approximately zero. \citet{2014ApJS..211...19B} list $\eta = 10^{-3}$ as the used standard value, consistent with usual truncation errors in the simulations. 

\subsubsection*{Gravity} 

Of particular importance for our planet formation model is the inclusion of the self-gravity of the gas. To achieve this, Poisson's equation 
\begin{align}
\nabla^2 \phi = 4 \pi G \rho, \label{eq:poisson2}
\end{align} 
with the gravitational potential $\phi$, the gravity constant $G$, and the gas density $\rho$, is solved in a multistep process for all cells. The cloud-in-cell (CIC) interpolation method from \cite{1988csup.book.....H} is used to approximate the gas distribution as a spatially-discretized density field, from which the gravitational potential is generated via a fast Fourier transform. To get to know the acceleration at each (sub)grid a finite difference scheme with Dirichlet boundary conditions is used and protected against oscillations through buffer zones surrounding the active grid zones. Because of approximations in this process, \citet{2014ApJS..211...19B} estimates the resolution of the gravitational force to be approximately twice as coarse as the corresponding refinement level.

\subsubsection*{Equation of state}

The total fluid energy density is given by
\begin{align}
E = e + \frac{\rho v^2}{2},\label{eq:totE}
\end{align}
with the thermal energy density $e$ governed by the equation of state
\begin{align}
e = \frac{p}{\gamma - 1}.
\end{align}
This is the equation of state for an ideal gas with the ratio of specific heats $\gamma$.

\subsubsection*{Computational domain} 

To prevent interactions of the disk material with the boundaries of the computational domain, we choose the physical size of the box to be (R$_{disk}\cdot 10)^3$. Thus, there is enough space for the disk to evolve and to be embedded in a diffuse medium. The boundaries of the domain behave as reflecting walls, i.e., 
\begin{align}
q (-x) = q(+x),
\end{align}
with an arbitrary field quantity $q$, and with the velocity perpendicular to the boundary direction reversed
\begin{align}
v_x (-x) = - v_x (+x).
\end{align}

\subsubsection*{Refinement criteria}
The refinement to the next (deeper) level of the grid patches in our simulations depends on the corresponding cell mass and a so-called Jeans length criterion. The cell mass refinement works such that it mimics a Lagrangian method in trying to keep a fixed mass resolution. Thus, if the mass of a specific cell is
\begin{align}
M_g = \rho (\Delta x)^d > \rho_{flag} (\Delta x_{root})^d r^{\epsilon_l l} = M_{\text{Jupiter}}/1000,
\end{align}
the cell is refined to a deeper level (if allowed by the maximum refinement level).
Here, $\rho_{flag}$ is the required density on the root grid, $\Delta x_{root}$ is the root grid cell spacing, $r$ the refinement factor (usually 2), $l$ the level and $\epsilon_l$ is an aggressiveness parameter, to make the refinement super-Lagrangian ($\epsilon_l < 0$) or sub-Lagrangian ($\epsilon_l > 0$). 

The second approach refines the Jeans length
\begin{align}
\lambda_J = \left( \frac{\pi c_s^2}{G \rho} \right)^{1/2}
\end{align}
by a fixed number of cells \citep[based on][]{1998ApJ...495..821T} if
\begin{align}
\Delta x < \left( \frac{\pi k_B T}{N_J^2 G \rho m_H} \right)^{1/2},
\end{align}
with $N_J = 32$ the required number of cells per Jeans length. This is especially valuable since we deal with fragmentation effects, where contraction plays a key role in the later building of clumps.

\subsection{Initial conditions}
\label{sec:IC}

\subsubsection*{3D density profile} 
To begin with, we set the column density structure of the disk according to a classical Mestel power-law profile \citep{1963MNRAS.126..553M}
\begin{align}
\Sigma (r) = \Sigma_0 \frac{r_{\text{out}}}{r},
\end{align}
with $r_{\text{out}}$ denoting the outer radius of the disk and $\Sigma_0$ the column density at this radius. Since the contribution to mass and angular momentum at small disk radii can be neglected we assume that the disk extends to $r = 0$. This leads to a normalization of the column density as
\begin{align}
\Sigma_0 = \frac{1}{2 \pi} \frac{M_{\text{disk}}}{r_{\text{out}}^2}, 
\end{align}
with the overall disk mass chosen to be $M_{\text{disk}} \approx 0.05 \, M_{\odot}$. Using hydrostatic balance the analytical solution to resolve the Gaussian density distribution in the vertical direction is
\begin{align}
\rho (r,z) = \frac{\Sigma (r)}{H(r) \sqrt{2 \pi}} \text{exp}\left(- \frac{z^2}{2 H(r)^2} \right). \label{eq:density}
\end{align}
Here, $z$ denotes the distance to the midplane of the disk and $H$ the scale height, which is evaluated as \citep{2008NewAR..52...21L}
\begin{align}
H(r) & = \frac{H_{\text{sg}}}{2} \left( \frac{H_{\star}}{H_{\text{sg}}} \right)^2 \left[ \sqrt{1+4\left(\frac{H_{\text{sg}}}{H_{\star}}\right)} -1 \right] \approx 0.62 H_{\star} , \label{eq:H}
\end{align}
with the scale height for self-gravitating disks 
\begin{align}
H_{\text{sg}} = c_s^2 / \pi G \Sigma,
\end{align} 
and the scale height for Keplerian disks
\begin{align}
H_{\star} = c_s / \Omega. \label{eq:Hstar}
\end{align}
In the equation above, the sound speed is given via the Newton-Laplace equation
\begin{align}
c_s = \sqrt{\gamma \cdot \frac{p}{\rho}} \label{eq:NL},
\end{align}
and the angular velocity $\Omega$ (see next subitem). As mentioned by \cite{2008NewAR..52...21L}, this approximation holds for a comparable gravitational influence of the disk and the central object, $H_{\text{sg}} \approx H_{\star}$.

\subsubsection*{Central object} 
We choose the host star of the system to be of subsolar mass 
\begin{align}
M_{\star} = 0.646 M_{\odot},
\end{align}
which is represented in the simulations as a point mass particle without spatial extension (denoted in the later figures with a single black dot in the middle) and is included in the gravity calculations.

\subsubsection*{Orbital velocity profile} 
Including the disk's self-gravity, the azimuthal velocity satisfies \citep{2008NewAR..52...21L}
\begin{align}
\frac{v_{\phi}^2}{r} & = \frac{\partial \Phi_{tot}}{\partial r} + \frac{1}{\rho} \frac{\mathrm{d} p}{\mathrm{d} r}
 = \frac{G M_{\star}}{r^2} + 2 \pi G \Sigma + \frac{1}{\rho} \frac{\mathrm{d} p}{\mathrm{d} r},
\end{align}
with $\Phi_{tot}$ the gravitational potential of central object and disk material.

\subsubsection*{Thermal profile} 
Finally, we need to address the thermal profile of the disk. We characterize the disk through the Toomre parameter \citep{1964ApJ...139.1217T}
\begin{align}
Q = \frac{c_s \kappa}{\pi G \Sigma} \approx \frac{c_s \Omega}{\pi G \Sigma}, \label{eq:ToomreQ}
\end{align}
with Boltzmann's constant $k_B$. Fixing $Q$ to a constant value then translates into a condition for the sound speed and therefore the temperature of the gas. The epicyclic frequency $\kappa$ can be approximated with the angular velocity $\Omega = v_{\phi}/r$, when the velocity profile throughout the disk is mostly dominated by the Keplerian velocity \citep{2008NewAR..52...21L}. We assume a gas with composition similar to the solar system, i.e., the mean molecular weight $\mu$ is set to $2.4$ \citep{2007ApJ...661L..77M}.
From this assumption we infer the temperature profile using Equation \ref{eq:NL} to be
\begin{align}
T(r) = \frac{c_s^2 \mu m_p}{\gamma k_B}. \label{eq:temperature}
\end{align}
The ratio of specific heats is fixed to be $\gamma =1.0001 \approx 1$, according to an isothermal equation of state (EOS).

\section{Resolution study}
\label{sec:resolution}

In this chapter, we investigate the imprint of the refinement level on the physical conditions within the disk. We benchmark the simulations for disks with $R_{disk} = 100$ AU for the initial conditions and at $t = 1.5 \, T_{disk}$, i.e., 1.5 orbital times of the most outer radius of the disk. An overview of all of our featured simulations is given below in Table \ref{tab:runs}.

\begin{table}[!h]
\centering
\begin{tabular}{lcccc}
Run&$R_{disk}$ [AU]&$Q_{init}$&$g_{c}$&$l_{max}$\\
\hline
\texttt{R10i64r2} & 10 & 1 & $64^3$ & 2\\
\texttt{R10i64r3} & 10 & 1 & $64^3$ & 3\\
\texttt{R10i64r4} & 10 & 1 & $64^3$ & 4\\
\texttt{R10i128r2} & 10 & 1 & $128^3$ & 2\\
\texttt{R10sg256} & 10 & 1 & $256^3$ & 0\\
\\
\texttt{R100i64r2} & 100 & 1 & $64^3$ & 2\\
\texttt{R100i64r3} & 100 & 1 & $64^3$ & 3\\
\texttt{R100i64r4} & 100 & 1 & $64^3$ & 4\\
\texttt{R100i128r2} & 100 & 1 & $128^3$ & 2\\
\texttt{R100sg256} & 100 & 1 & $256^3$ & 0\\
\\
\texttt{R300i64r2} & 300 & 1 & $64^3$ & 2\\
\texttt{R300i64r3} & 300 & 1 & $64^3$ & 3\\
\texttt{R300i64r4} & 300 & 1 & $64^3$ & 4\\
\texttt{R300i128r2} & 300 & 1 & $128^3$ & 2\\
\texttt{R300sg256} & 300 & 1 & $256^3$ & 0\\
\\
\texttt{Qi64r2} & 100 & 3 & $64^3$ & 2\\
\texttt{Q3i64r3} & 100 & 3 & $64^3$ & 3\\
\texttt{Q3i64r4} & 100 & 3 & $64^3$ & 4\\
\texttt{Q3i128r2} & 100 & 3 & $128^3$ & 2\\
\texttt{Q3sg256} & 100 & 3 & $256^3$ & 0\\
\end{tabular}
\caption{\label{tab:runs} Overview of the settings of all performed simulation runs, with the disk radius $R_{disk}$, the Toomre parameter $Q_{init}$, the initial number of cells of the coarsest grid in the simulation box $g_{c}$, and the maximum allowed refinement level $l_{max}$.}
\end{table}

\begin{figure}[!t]
        \centering
        \begin{subfigure}[b]{0.5\textwidth}
                \includegraphics[width=\textwidth]{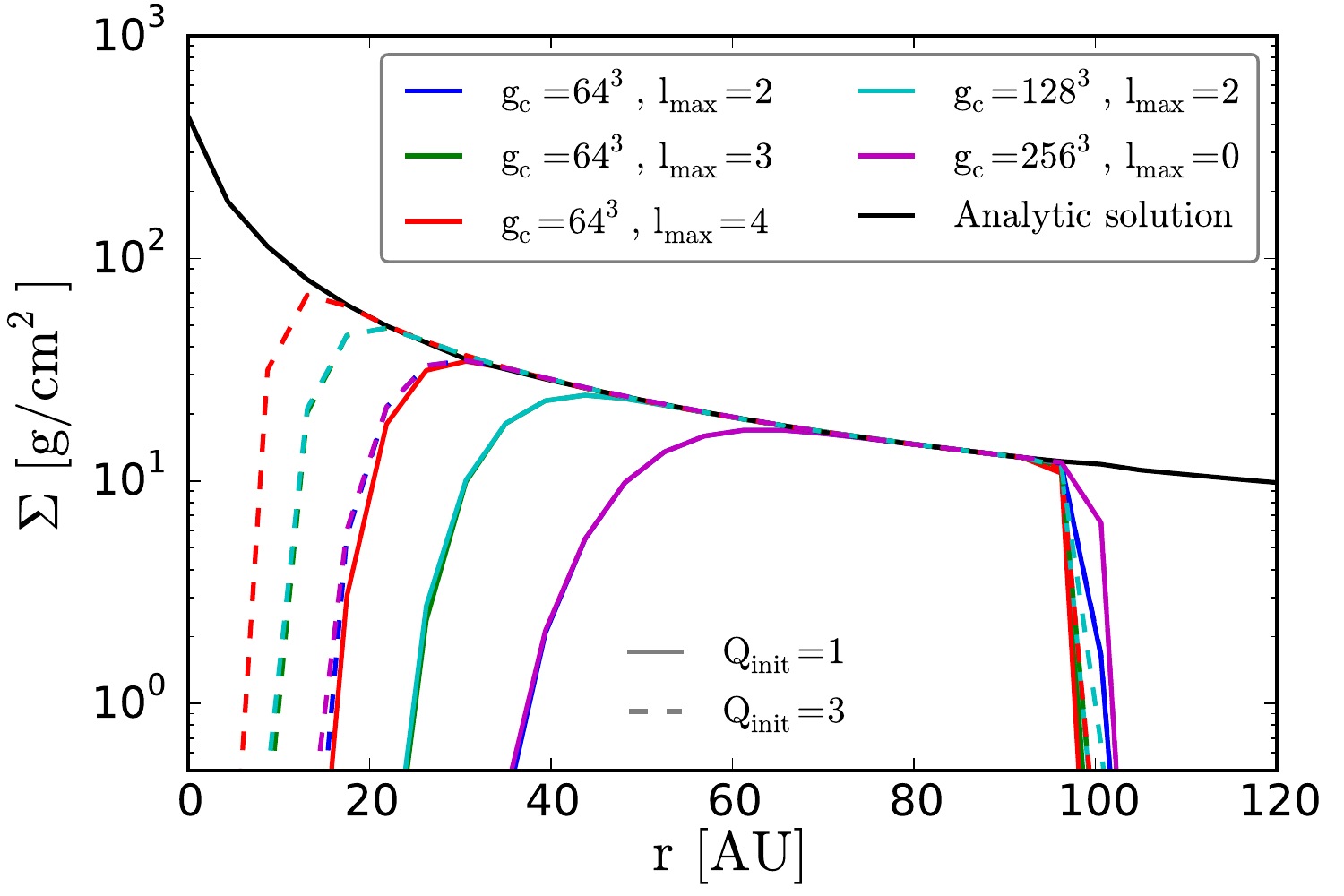}
        \end{subfigure} \\
        \begin{subfigure}[b]{0.5\textwidth}
                \includegraphics[width=\textwidth]{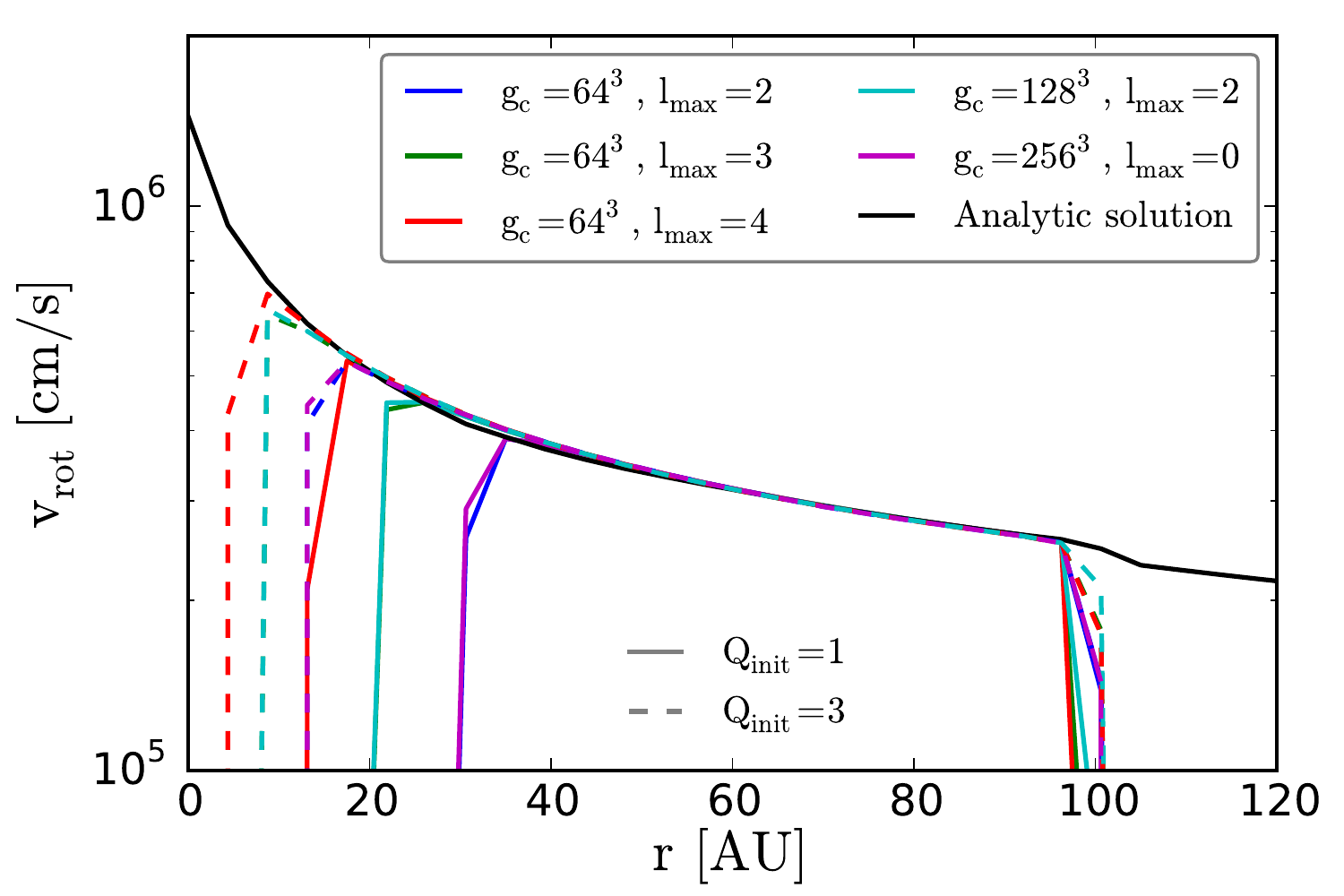}
        \end{subfigure} 
        \caption{Radial profiles for initial column density $\Sigma$ and rotational velocity $v_{rot}$ for simulations with $R_{disk} = 100$. The parameter $g_c$ indicates the resolution of the coarsest grid, $l_{max}$ the deepest level of the resolution by dynamic refinement. The continous lines show simulations with $Q_{init}=1$, the dashed lines simulations with $Q_{init}=3$. The density distribution converges toward the analytic solution for higher refinement levels, i.e., the density gap in the inner part of the disk is smaller for higher refinement levels. Additionally, it is smaller for the simulation runs with $Q_{init} = 1$, which is due to the scale height dependency on the $Q$ parameter (compare Equation \ref{eq:ToomreQ} and Figure \ref{fig:DensitySliceZ_T0}). The velocity, only nonzero within the disk matter, behaves analogous to the density.}\label{fig:IC_DensityVelocity}
\end{figure}

\begin{figure}[!t]
        \centering
        \begin{subfigure}[b]{0.5\textwidth}
                \includegraphics[width=\textwidth]{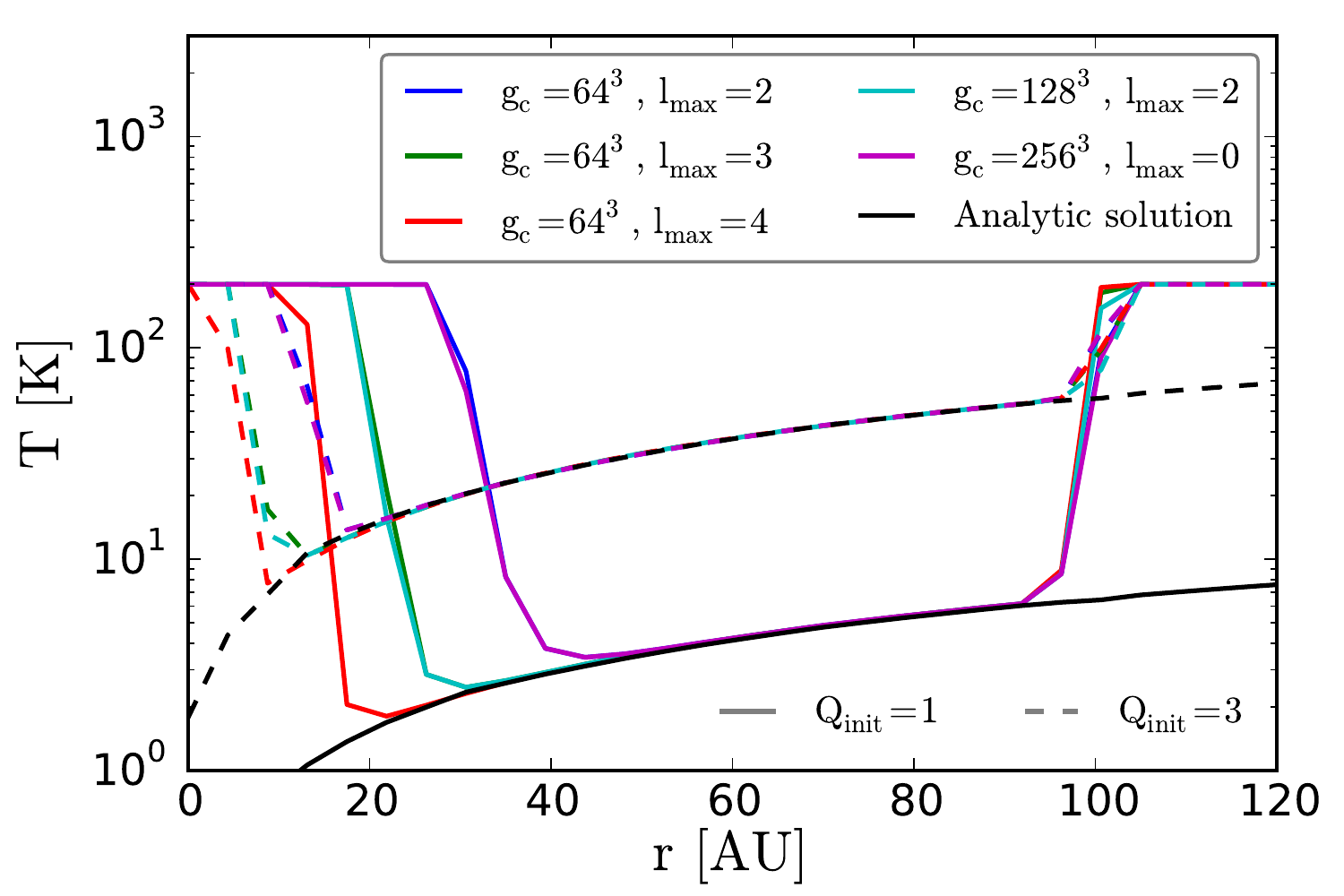}
        \end{subfigure} \\
        \begin{subfigure}[b]{0.5\textwidth}
                \includegraphics[width=\textwidth]{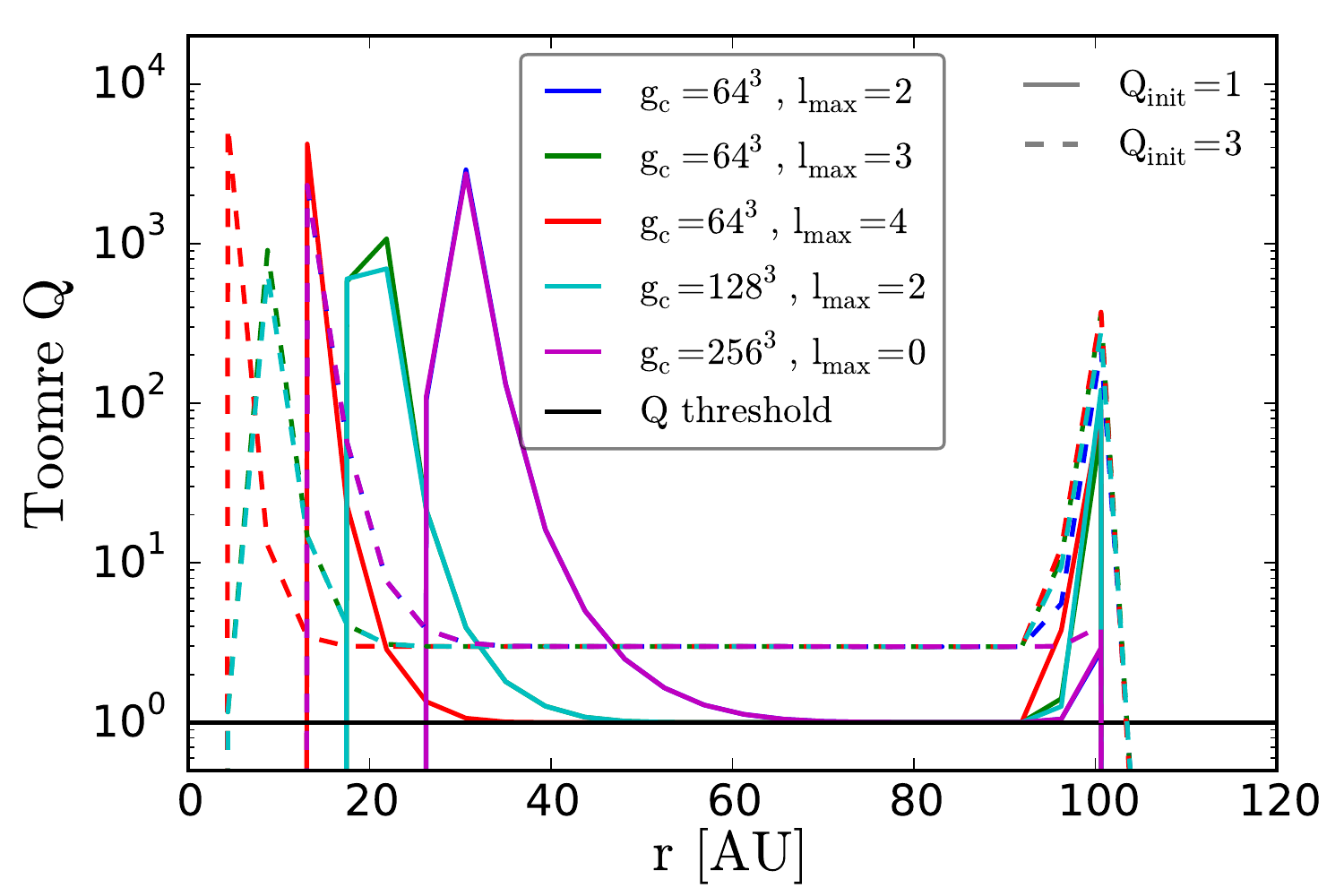}
        \end{subfigure} 
        \caption{Radial profiles for initial temperature $T$ and effective Toomre $Q$ parameter for simulations with $R_{disk} = 100$. The temperature increases outward because of its dependence on $Q$ (see Equation \ref{eq:temperature}). $Q$ is approximated correctly in the inner parts of the disk. In outer parts, where the density decreases for numerical reasons, $Q$ rises, and thus artificially stabilizes the disk matter.}\label{fig:IC_TemperatureToomreQ}
\end{figure}

\subsection{Initial state}
\label{sec:initial_state}

Figure \ref{fig:IC_DensityVelocity} and \ref{fig:IC_TemperatureToomreQ} show comparisons of the radial profiles of the values of $\Sigma, v_{rot}, T$ and $Q$ of disks with $R_{disk} = 100$ AU and $Q_{init} = 1$, $Q_{init} = 3$, respectively. All of them converge for higher maximum refinement levels toward the analytical solution, as given in Section \ref{sec:IC}.
Some differences between the simulations with $Q_{init}=1$ and $Q_{init}=3$ arise because of the change of scale height $H$ by a change in $Q$, according to Equation \ref{eq:Hstar}. 
It is important to say that the simulations with the same effective resolution behave very similarly, despite any differences in resolution for more coarse grids. Therefore, we restrict the discussion on the simulations with $g_{c} = 64^3$, while analogous simulations with the same effective resolution can be identified for the runs with $g_c=128^3$. In the following we discuss and explain the differences in detail.

\begin{figure*}[bth]
        \centering
        \includegraphics[width=\textwidth]{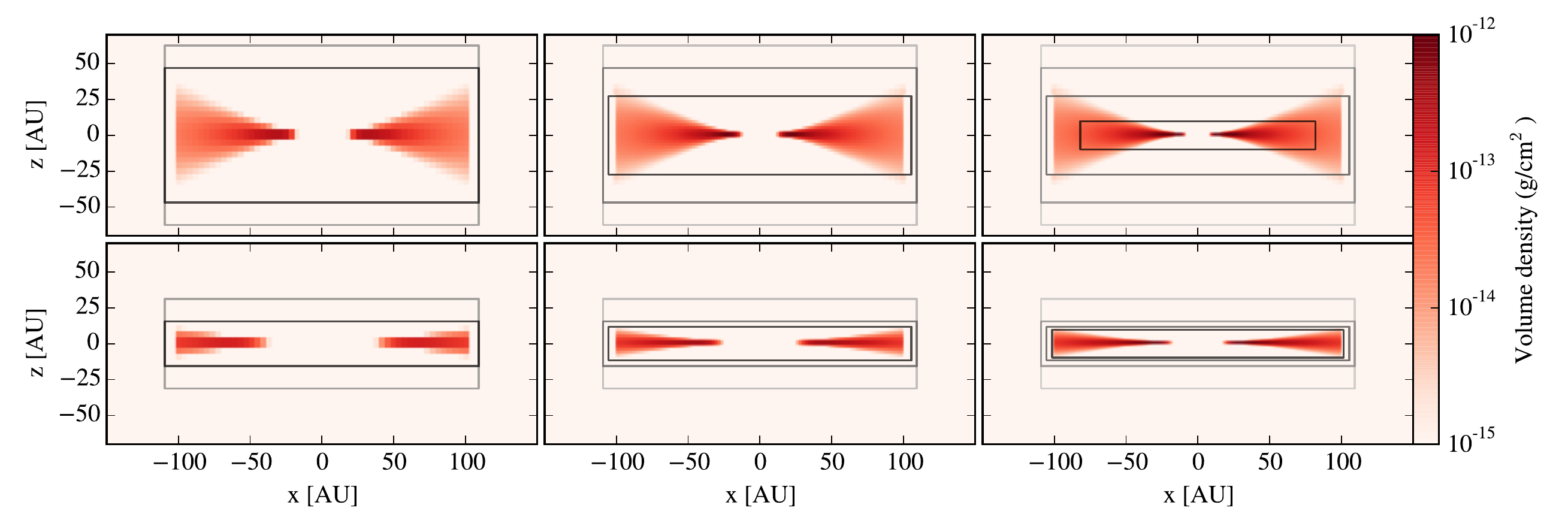}
        \caption{The initial vertical density profile $\Sigma$, comparing disks with $R_{disk} = 100$ AU and $Q_{init} = 3$ (top), $Q_{init} = 1$ (bottom), and increasing maximum refinement level from 2 - 4 (left to right), respectively. The scale height $H$, according to Equation \ref{eq:Hstar}, changes with Q, which is why the $Q_{init}=1$ disks show a flatter profile. The gray boxes indicate the deepness of the refinement level, from level $l = 0$ outside of the boxes to level $l = l_{max}$ in the most inner box.}\label{fig:DensitySliceZ_T0}
\end{figure*}

The profiles for the surface density $\Sigma$ yield two conclusions. First, higher refinement means that the disk is much better resolved, especially in the inner part. Second, it displays the dependency of $H$ on $Q$. This is illustrated in Figure \ref{fig:DensitySliceZ_T0}, which shows slices of the initial density profile in a cut through the vertical profile of the disk, comparing disks with $R_{disk} = 100$ AU and $Q_{init} = 1$, $Q_{init} = 3$, respectively. The different refinement levels are indicated with gray boxes. These show that the disk midplane is better resolved than the outer parts, thus enabling a higher resolution in the regimes important for fragmentation. The vertical disk extension is no longer resolved when the scale height drops below the cell size at small radii.

All disks represent the analytical solution for the rotational velocity very well, with only minor differences. They are better resolved (i.e., better converge toward the analytic solution in the inner regions) for a higher refinement level. This is a direct consequence from the density distribution (figures \ref{fig:IC_DensityVelocity} and \ref{fig:DensitySliceZ_T0}), since the rotational velocity is only defined in the disk material.

The temperature profiles show fundamental differences for the $Q_{init}=3$ and $Q_{init}=1$ cases as a result of the $Q$ dependence of the thermal profile (Equation \ref{eq:temperature}). Fixing the Toomre parameter to $Q_{init}=1$ results in lower temperatures throughout the disk and thus an increased chance for developing fragmentation. As a result, the temperature distributions show a positive slope with increasing radius, which is adopted to explore disks in a marginally stable state. Outside of the disk the temperature rapidly increases to the temperature of the background gas $\mathrm{T}_{background} = 200$ K.

Finally, the effective Toomre $Q$ distributions show the numerical imprint of $Q_{init}$ on the disk as a consequence of discretization effects. Thus, $Q$ drastically increases in the inner and outer parts of the disk, where the density per cell decreases for numerical reasons. Despite these limitations $Q$ stays constant in the dynamically relevant disk parts, where the density is high enough to support fragmentation effects.

\subsection{Evolved state}
\label{sec:evolved}

\begin{figure}[bth]
        \centering
        \begin{subfigure}[b]{0.5\textwidth}
                \includegraphics[width=\textwidth]{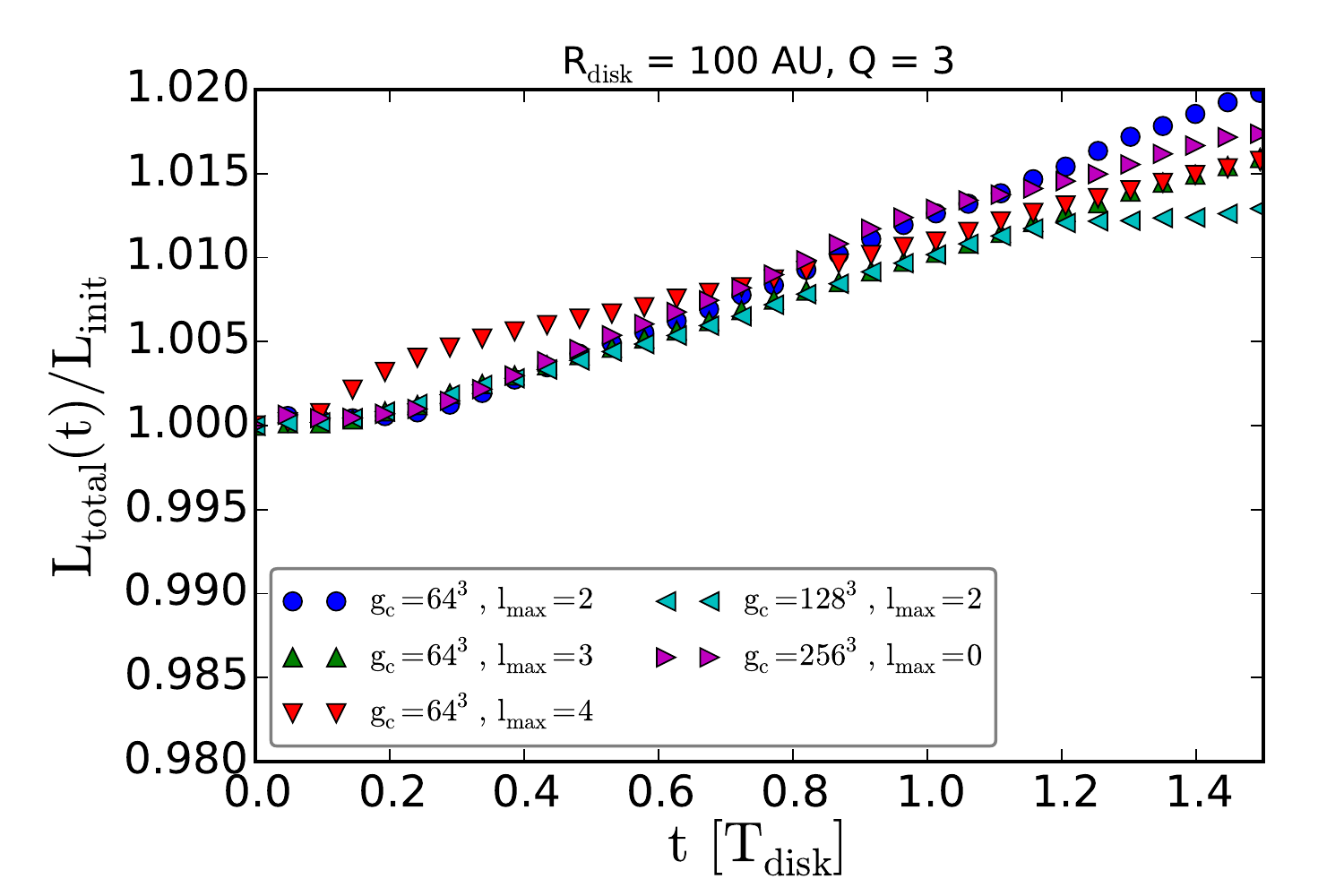}
        \end{subfigure} \\
        \begin{subfigure}[b]{0.5\textwidth}
                \includegraphics[width=\textwidth]{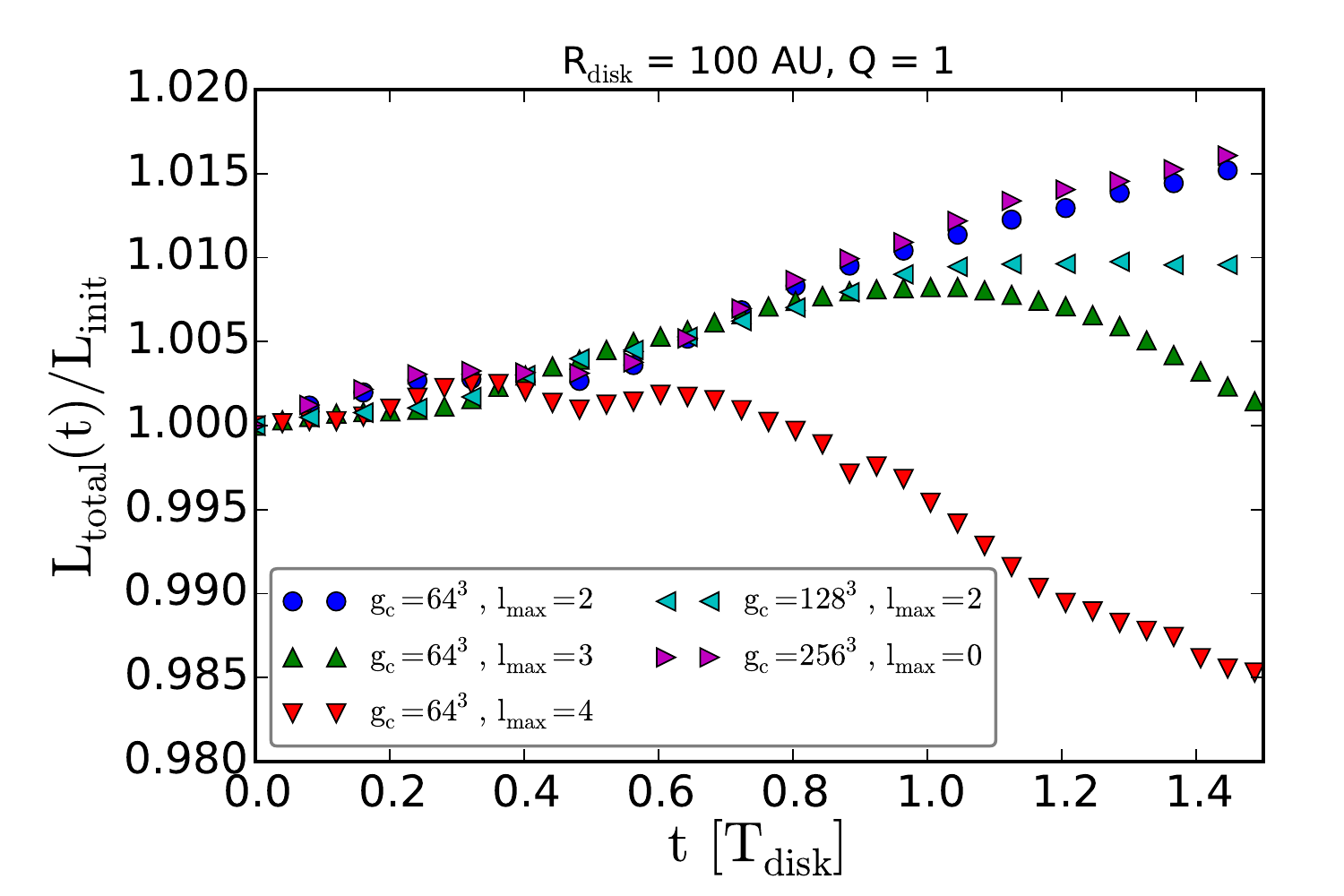}
        \end{subfigure}
        \caption{Evolution of angular momentum for all simulations with $R_{disk} = 100$ and up to $t = 1.5 \, T_{disk}$. The maximum deviations from the initial state reach up to 2\%.}\label{fig:TotalAM}
\end{figure}

A potential weakness of Cartesian grid codes is angular momentum dissipation, which is especially relevant for modeling systems with high rotational velocities. Therefore, to demonstrate the numerical robustness of our setup, we show the evolution of the total angular momentum $L_{total}$ in the whole simulation box in Figure \ref{fig:TotalAM} for simulation times up to $t = 1.5 \, T_{disk}$. 

\begin{figure*}[tbh]
        \centering
        \includegraphics[width=\textwidth]{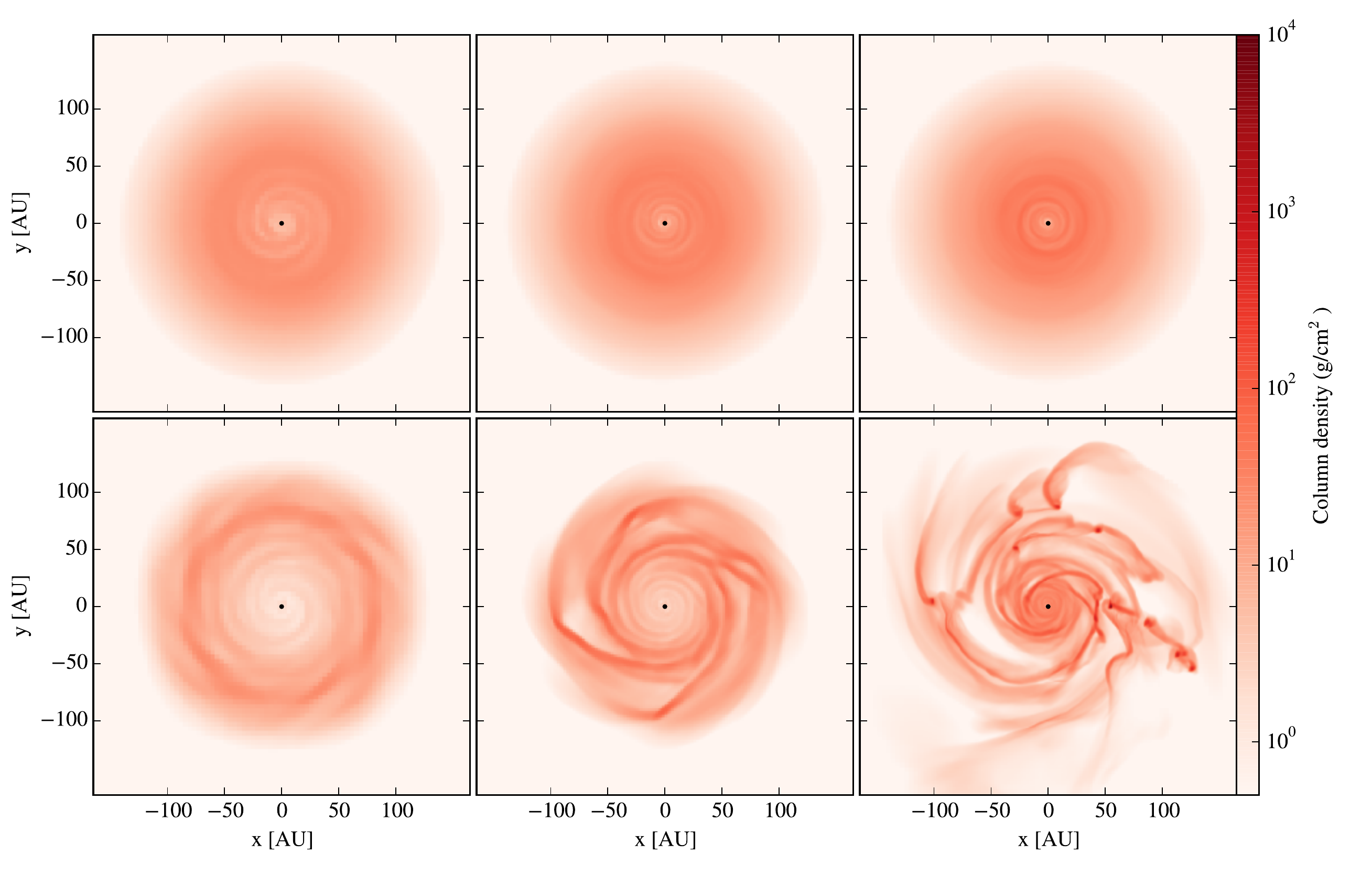}
        \caption{Face-on column density profiles for simulations with $g_{c} = 64^3$ and $t = 1.5 \, T_{disk}$. The top row shows simulations with $Q_{init}=3$, the bottom row simulations with $Q_{init}=1$. The maximum refinement level $l_{max}$ is increasing from left to right from 2 - 4.}\label{fig:Q3_prj}
\end{figure*}

\begin{figure*}[tbh]
        \centering
        \includegraphics[width=\textwidth]{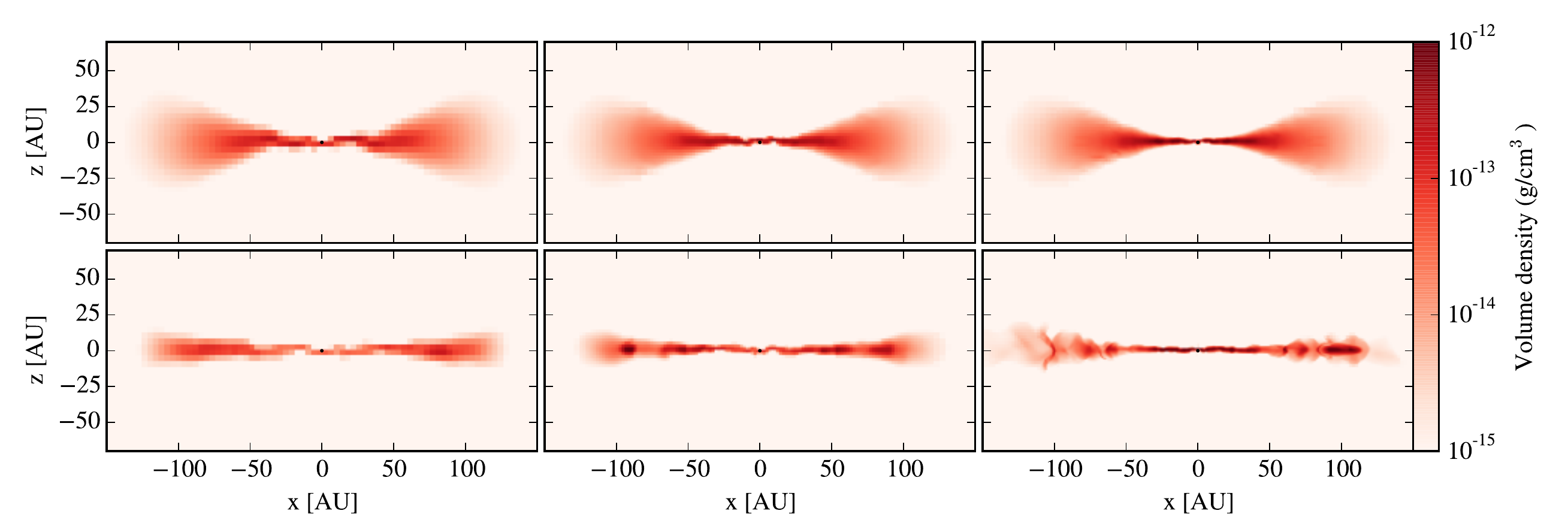}
        \caption{Edge-on volume density profiles for simulations with $g_{c} = 64^3$ and $t = 1.5 \, T_{disk}$. The top row shows simulations with $Q_{init}=3$, the bottom row simulations with $Q_{init}=1$. The maximum refinement level $l_{max}$ is increasing from left to right from 2 - 4.}\label{fig:Evolved_sliceZ}
\end{figure*}

Overall, the deviations for all simulations go at maximum up to +2\% of the initial angular momentum. For the simulations with $Q_{init} = 3$, the angular momentum shows a very stable configuration and all simulations show the same general trend. However, the simulations with $Q_{init}=1$ differ from each other, depending on the deepest refinement level in the simulation. Here, the simulations with the highest resolutions (i.e., with deepest refinement level $l_{max} =3$ or $l_{max} =4$) show a decrease of up to $-1.5$\% total angular momentum. 

\begin{figure}[tbh]
        \centering
        \begin{subfigure}[b]{0.5\textwidth}
                \includegraphics[width=\textwidth]{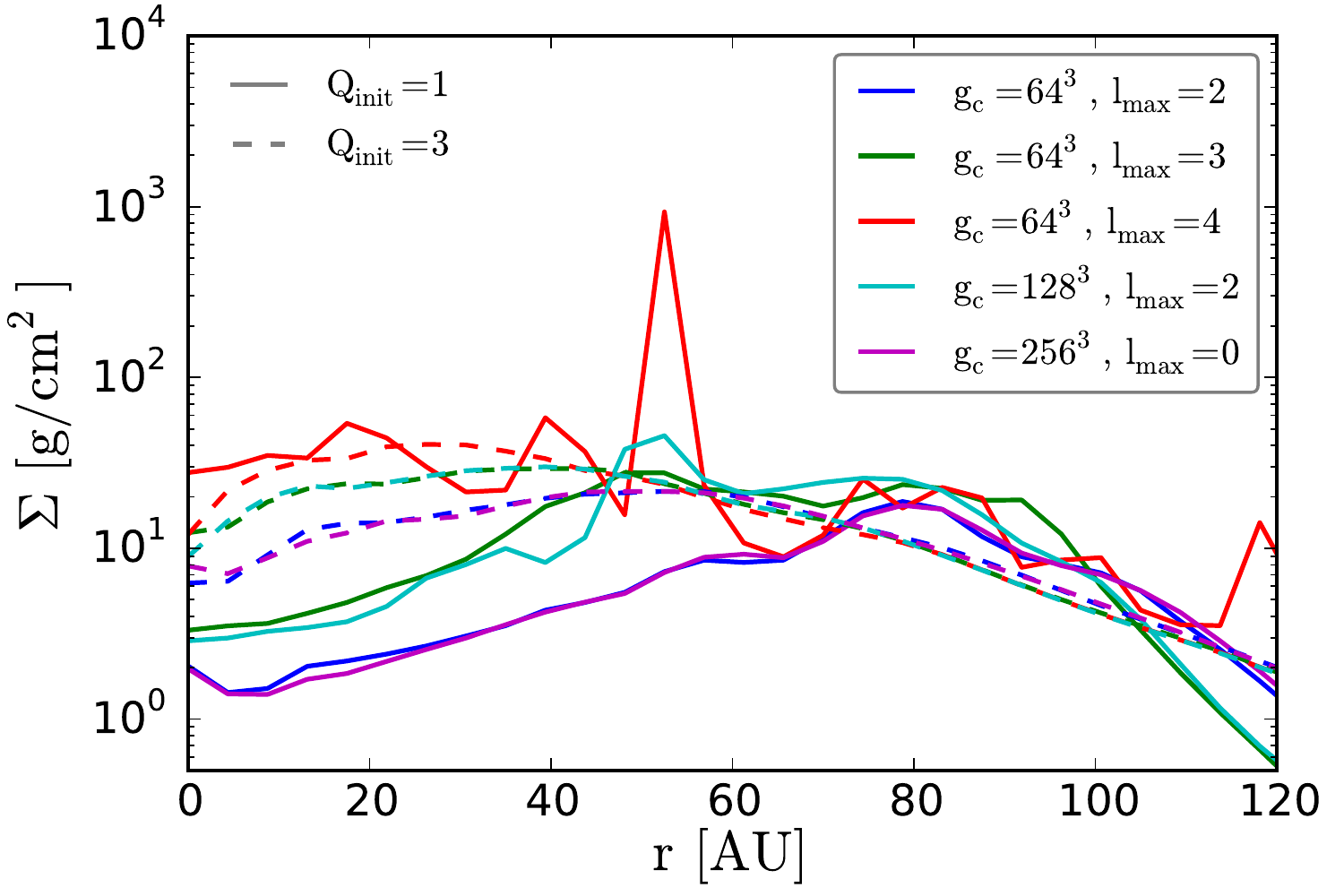}
        \end{subfigure} \\
        \begin{subfigure}[b]{0.5\textwidth}
                \includegraphics[width=\textwidth]{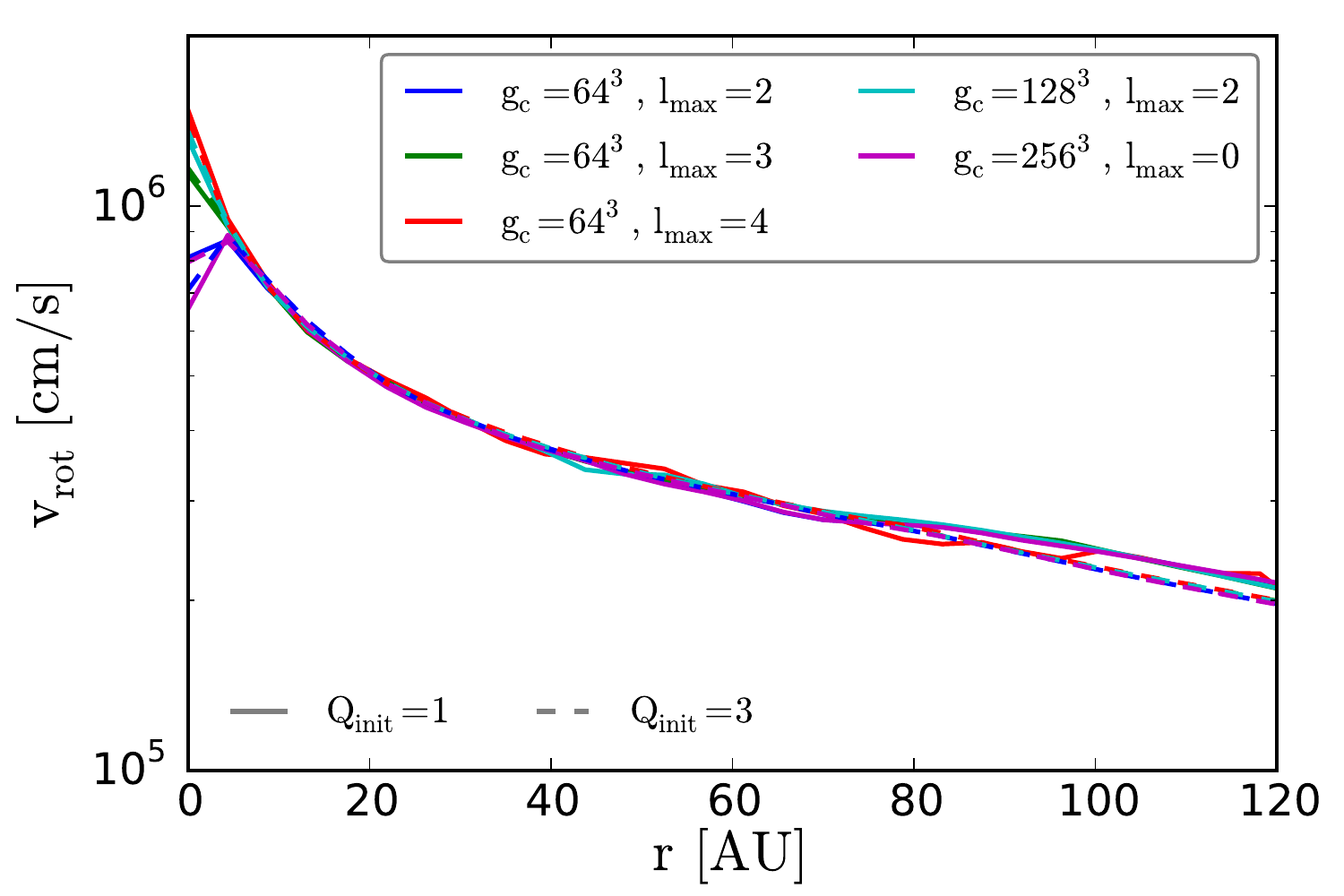}
        \end{subfigure}
        \caption{Radial profiles of column density $\Sigma$ and rotational velocity $v_{rot}$ for all simulations with $R_{disk} = 100$ and $t = 1.5 \, T_{disk}$. The runs with $Q_{init}=3$ approximately follow the initially chosen density distribution. The $Q_{init}=1$ disks show peaks at radii with distinct clumps or spiral structure (compare figures \ref{fig:Q3_prj} and \ref{fig:Evolved_sliceZ}). The velocity profiles show only minor deviations from the general trend, which are due to the turbluent velocity in the most unstable regions.}\label{fig:EvolvedRadProfile_SurfaceDensityVelocity}
\end{figure}

\begin{figure}[!t]
        \centering
        \begin{subfigure}[b]{0.5\textwidth}
                \includegraphics[width=\textwidth]{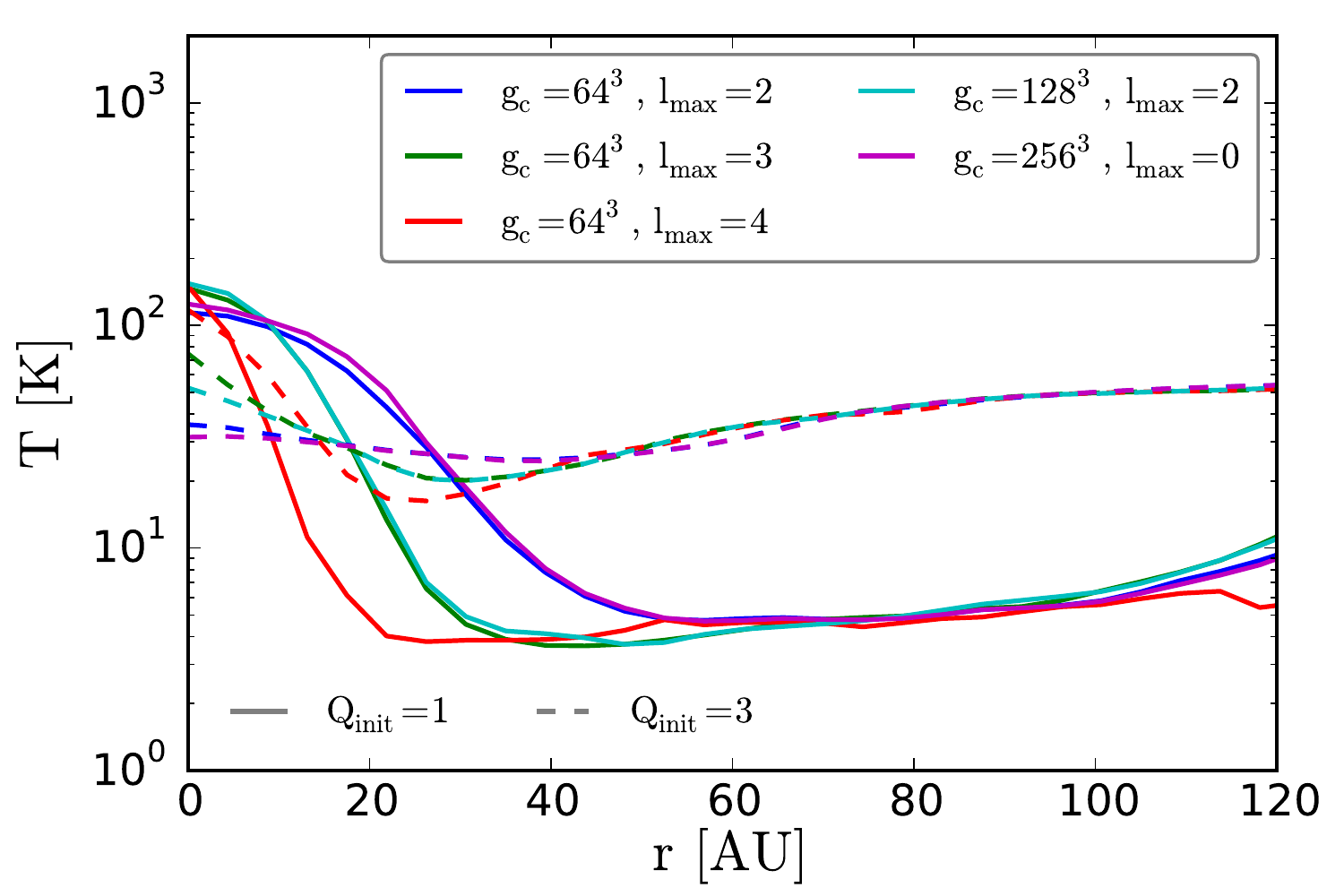}
        \end{subfigure} \\
        \begin{subfigure}[b]{0.5\textwidth}
                \includegraphics[width=\textwidth]{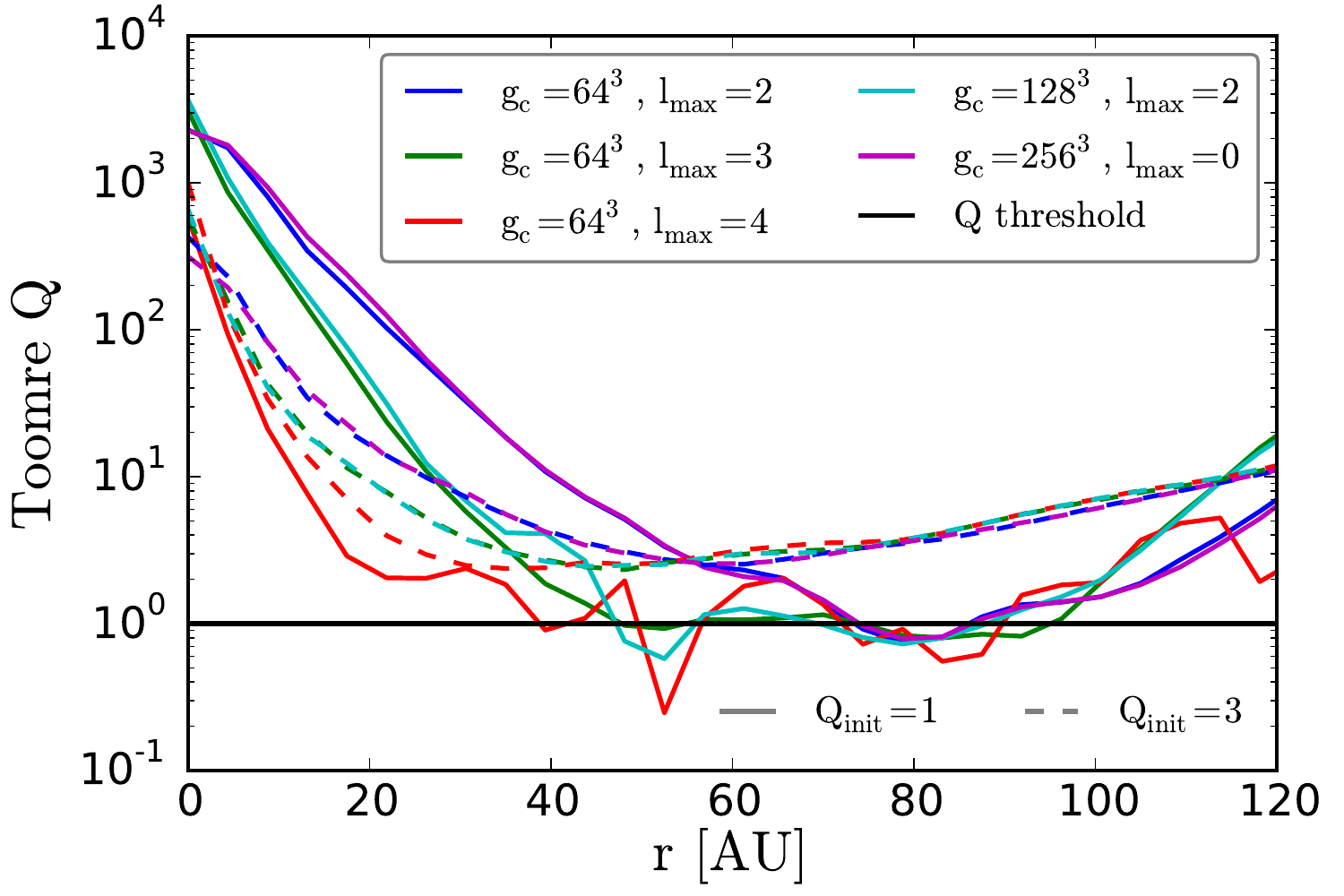}
        \end{subfigure}
        \caption{Radial profiles of the temperature distribution and effective Toomre $Q$ parameter for all simulations with $R_{disk} = 100$ and $t = 1.5 \, T_{disk}$. As the density distribution the temperature shows a spread-out for the inner and outer parts. This is much more pronounced for the simulations with $Q_{init}=1$, whereas the runs with $Q_{init}=3$ roughly follow the initial distribution. For $Q_{init}=3$, the effective $Q$ parameter stays well above the threshold for marginal stability. In comparison, the effective $Q$ in the $Q_{init}=1$ runs dips below the threshold at the positions with increased density (i.e., fragmenting regions, compare Figure \ref{fig:EvolvedRadProfile_SurfaceDensityVelocity}).}\label{fig:EvolvedRadProfile_TemperatureToomreQ}
\end{figure}

Now we focus on the imprint of changes in resolution on the evolved state of the disks for $t = 1.5 \, T_{disk}$ to spot deviations for different refinement depths. Figure \ref{fig:Q3_prj} visualizes the density distribution for a face-on view of the disks for various refinements for $Q_{init}=3$ and $Q_{init}=1$. The disks with $Q_{init}=3$ do not differ qualitatively with increasing refinement level, but show increased resolution and details in the gas streams.

However, the differences for the $Q_{init}=1$ disks are dramatic. Whereas the disk with $l_{max} = 2$ does not show any signs of clumping or spiral arms, these characteristics start to appear for $l_{max} = 3$ and are sharply defined in the simulation with $l_{max} = 4$. The differences due to $l_{max}$ do not only change the overall density distribution of the disk. In addition, the planetary (clump) wake, associated with each forming clump, differs dramatically from $l_{max} = 3$ to $l_{max} = 4$. For $l_{max} = 3$, the clump only marginally disturbs the large-scale structure of the spiral arms, forming regions of higher density at approximately 50 and 80-90 AU separation. In contrast, the simulation with $l_{max} = 4$ shows very distinct spots, where fragmentation occurs. These clumps massively disturb the spiral arm structure, caused by the turbulent motion in the fragmenting areas.

The edge-on views of the density distribution of the disks (Figure \ref{fig:Evolved_sliceZ}) reveal structural differences in the vertical gas distribution with increasing refinement level.
The $Q_{init}=3$ disks again show relatively weak differences, however, the inner part of the disk flattens with increasing refinement and thus the flared disk profile is better resolved.

The $Q_{init}=1$ disks show this flattening as well. Moreover, the density peaks in the midplane parts of the disks are more pronounced with higher refinement level. Comparing the edge-on slices of the disks with the face-on projections demonstrates the capabilities of three-dimensional simulations. In areas of strong fragmentation the disk's vertical extension is lower compared to more stable regions. Thus, the flared profile of the disk is disturbed and shows strong deviations from vertical and axisymmetry.

The visual observations are supported by the quantitative analysis of the radial profiles in figures \ref{fig:EvolvedRadProfile_SurfaceDensityVelocity} and \ref{fig:EvolvedRadProfile_TemperatureToomreQ}.

Figure \ref{fig:EvolvedRadProfile_SurfaceDensityVelocity} shows the density distributions of the evolved state. The simulations with $Q_{init}=3$ roughly follow the initally chosen density profile. In the inner and outer parts the profile is marginally spread out. The increase of density in the inner parts is the result of our analytic velocity approximation. Therein we assume the disk to be extended all the way down to the center. However, since the disk's inner extension is cut off at some radius (see Section \ref{sec:initial_state}) the pressure gradient at the inner radius of the density distribution is too high. Therefore, the effective velocities in the inner part of the numerical implementation are too small. Thus, some of the gas on lower orbits rapidly migrates toward the center until an equilibrium state is reached. A similar effect occurs in the outer parts, where the gas is spread out until a smooth transition from high to low density is reached.

In comparison, the density distributions of disks with $Q_{init}=1$ show deviations from the initially chosen profile. The turbulent and self-gravitating disk material induces stochastic fluctuations in the velocity. This, in turn, initiates the development of spiral arms and fragmentation, which is reflected in the radial profiles as peaks. Overall, the most massive clump can be seen for the simulation with $l_{max} = 4$ at \textasciitilde \, 52 AU, reaching column densities up to $10^3 \frac{g}{cm^2}$.

As discussed above, the velocity profiles for all disks reach an equilibrium state, which follows the analytical solution in most parts of the disk. The small deviations in the distribution for the $Q=1$ runs emerge from the spiral arms and fragmentation, which are more pronounced for higher maximum refinement.

The temperature distribution in Figure \ref{fig:EvolvedRadProfile_TemperatureToomreQ} behaves analogously to the density in Figure \ref{fig:EvolvedRadProfile_SurfaceDensityVelocity}. Thus, it features a spread in the inner and outer parts, more remarkable for lower resolutions. Higher resolutions however are able to sustain the initial profile better, which is true for both $Q_{init}=3$ and $Q_{init}=1$ simulations. All of them reflect the spread of the gas and feature an approximately similar temperature distribution to the outer edge.

The inital $Q$ state is well preserved for the $Q_{init}=3$ runs. However, the strong instabilities and the resulting fragmentation in the simulations with $Q_{init}=1$ lead to minima in the effective $Q$ below the unity threshold where the density reaches its highest values.

\begin{figure}[!t]
    \centering
            \includegraphics[width=0.5\textwidth]{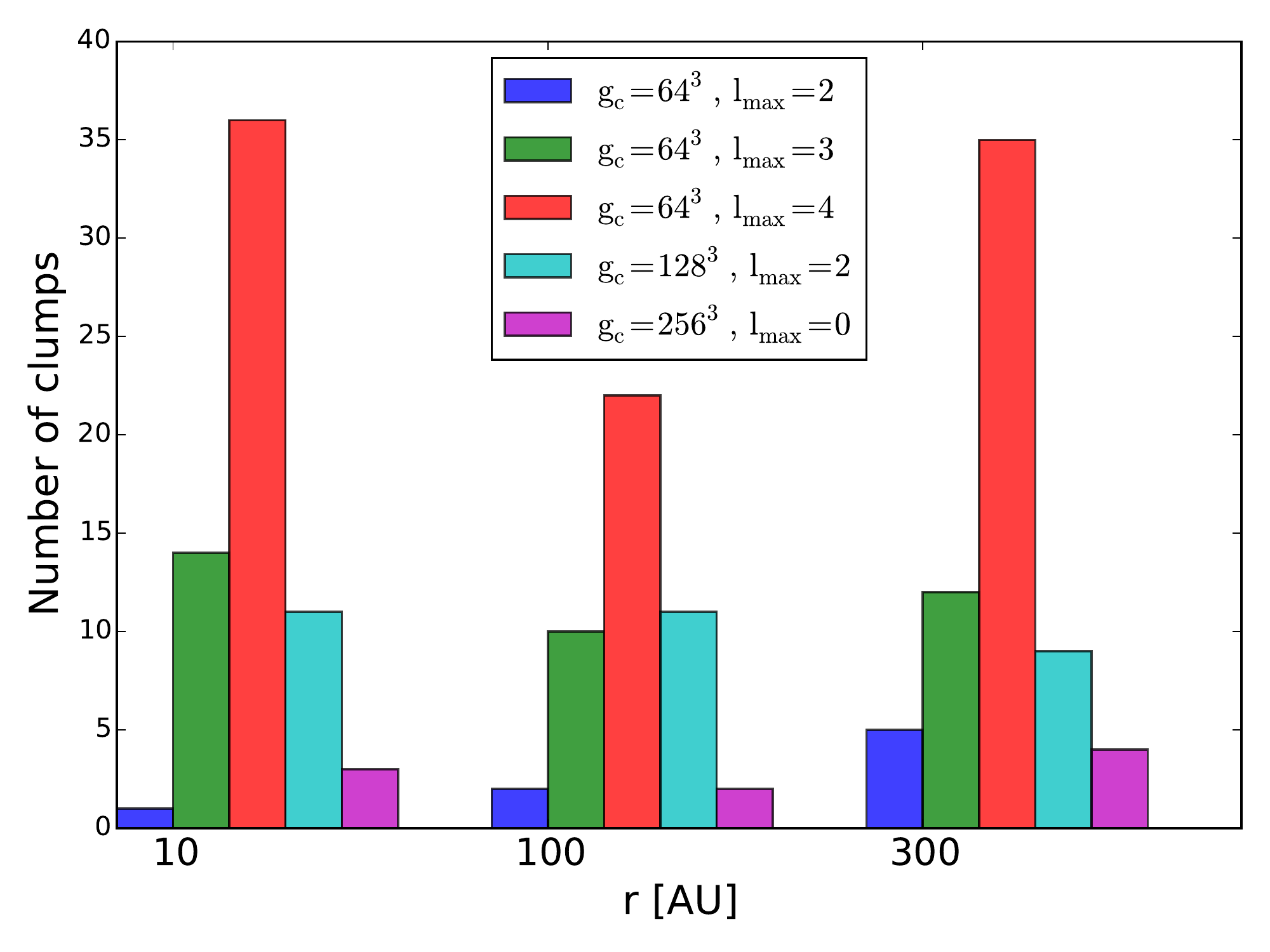}
    \caption{Number of lowest-level clumps for all simulations and $t = 1.5 \, T_{disk}$. The trend goes for just a few clumps for low maximum resolutions to many clumps for high maximum resolution. Different combinations of $g_{c}$ and $l_{max}$ but comparable resolution on the lowest level (i.e., green/cyan and blue/magenta) show approximately equal numbers of lowest-level clumps.}\label{fig:Clumps}
\end{figure}

As an attempt to quantify the behavior of the fragmentation for different refinement settings, we use the clump finding method of \citet{2010AIPC.1294..110S} to detect topologically disconnected structures within the dataset. Its principle mechanism is to create a single contour in between $\rho_{min}$ and $\rho_{max}$ over the whole computational domain. The lower value is then continously increased, until it reaches the maximum value. During that process disconnected structures are identified as separate contours and are treated as individual structures in which the routine continues recursively. To get an idea of the smallest structures in our simulations, we only print the far-end leaves (``clumps'') of this treelike structure, i.e., topologically disconnected regions featuring approximately the highest density in their specific contours. 

Figure \ref{fig:Clumps} illustrates our findings, showing a histogram with statistics for all simulations with $Q_{init} = 1$. The number of clumps increases with the maximum level of refinement, i.e., we find more clumps for higher resolution. 
This underlines the importance of a thorough resolution to achieve correct structures throughout the grid.

\section{Parameter study}
\label{sec:parameter}

\begin{figure}[!t]
    \centering
    \begin{subfigure}[b]{0.5\textwidth}
        \includegraphics[width=\textwidth]{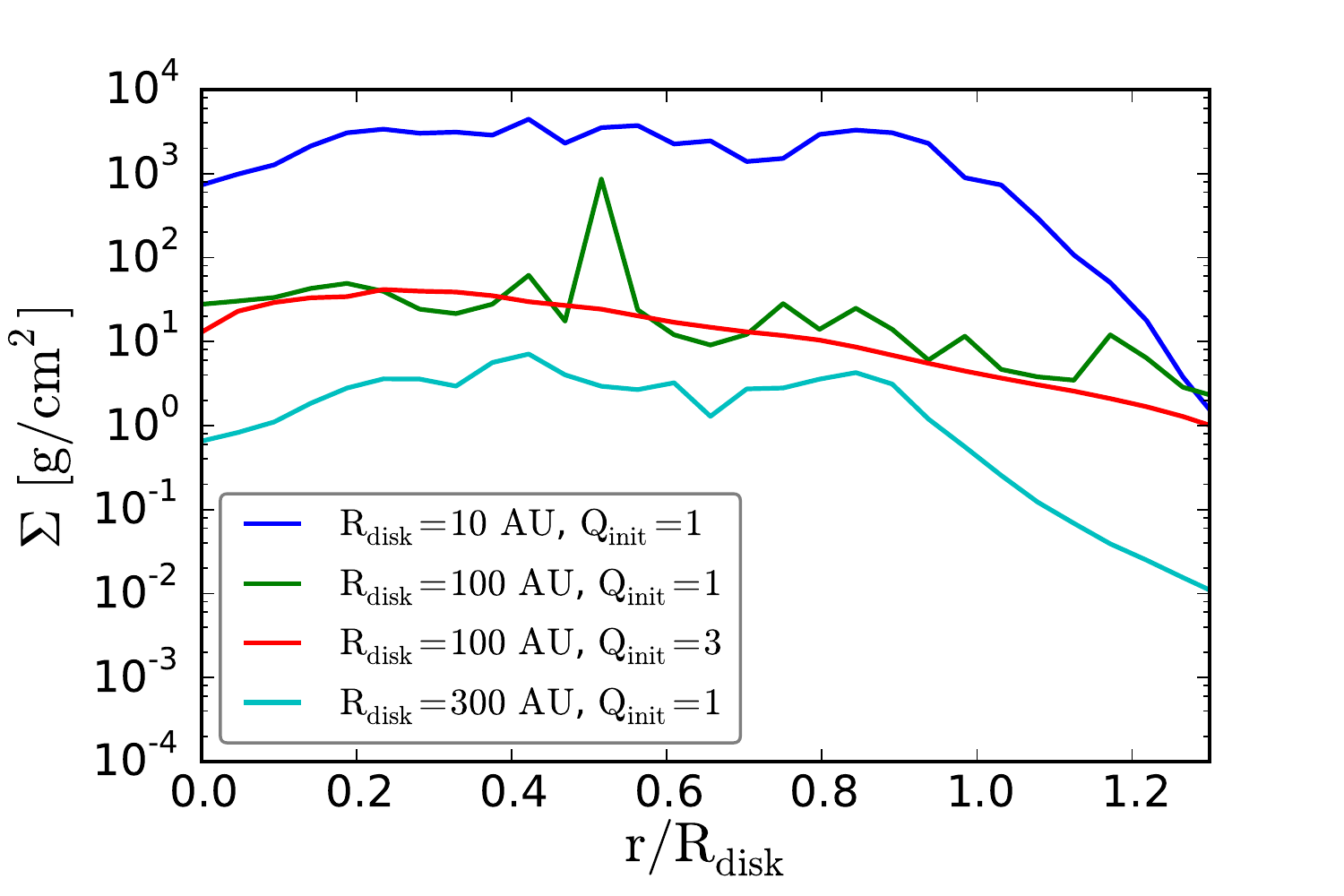}
    \end{subfigure} \\
    \begin{subfigure}[b]{0.5\textwidth}
        \includegraphics[width=\textwidth]{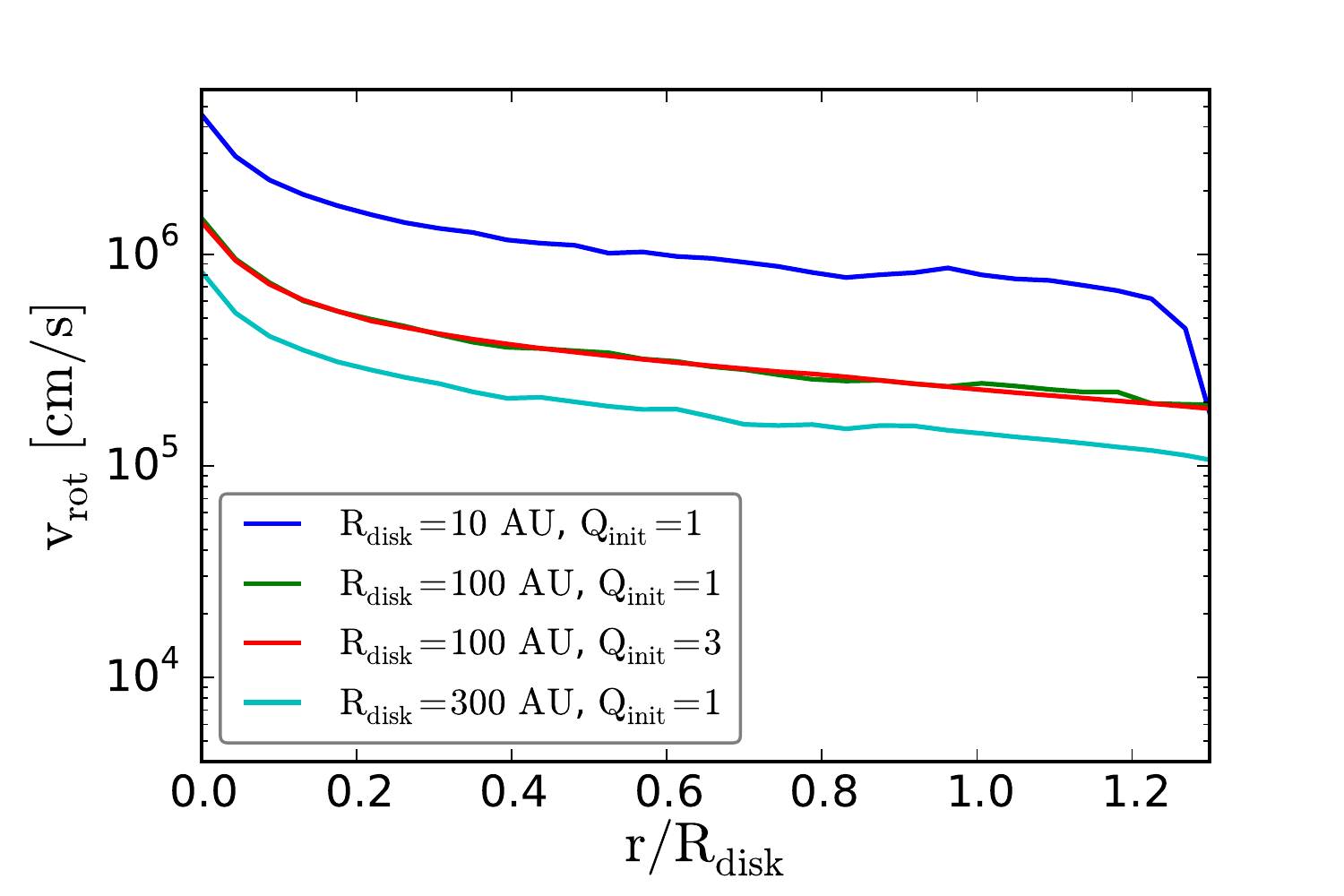}
    \end{subfigure} 
    \caption{Radial profiles at $t = 1.5 \, T_{disk}$ for column density $\Sigma$ and rotational velocity $v_{rot}$ for all simulations with the highest resolution on the finest grid, normalized in radial direction. The simulation with $Q_{init}=3$ is added for comparison. The density distributions for all radii show similarities. However, the simulation with $r=100$ AU and $Q_{init}=1$ shows its single peak at \textasciitilde $\, 52$ AU and the negative slope at the outer edge of the disk is much shallower than the slope of the other simulations with $Q_{init}=1$. The velocity distributions for all radii show a similar trend. The simulation with $r=10$ AU abruptly goes down in velocity. This behavior might be due to the very sharp transition from disk material (with approximately Keplerian velocity profile) to surrounding material (with approximately zero velocity).}\label{fig:PS_DensityVelocity}
\end{figure}

\begin{figure}[tbh]
    \centering
    \begin{subfigure}[b]{0.5\textwidth}
        \includegraphics[width=\textwidth]{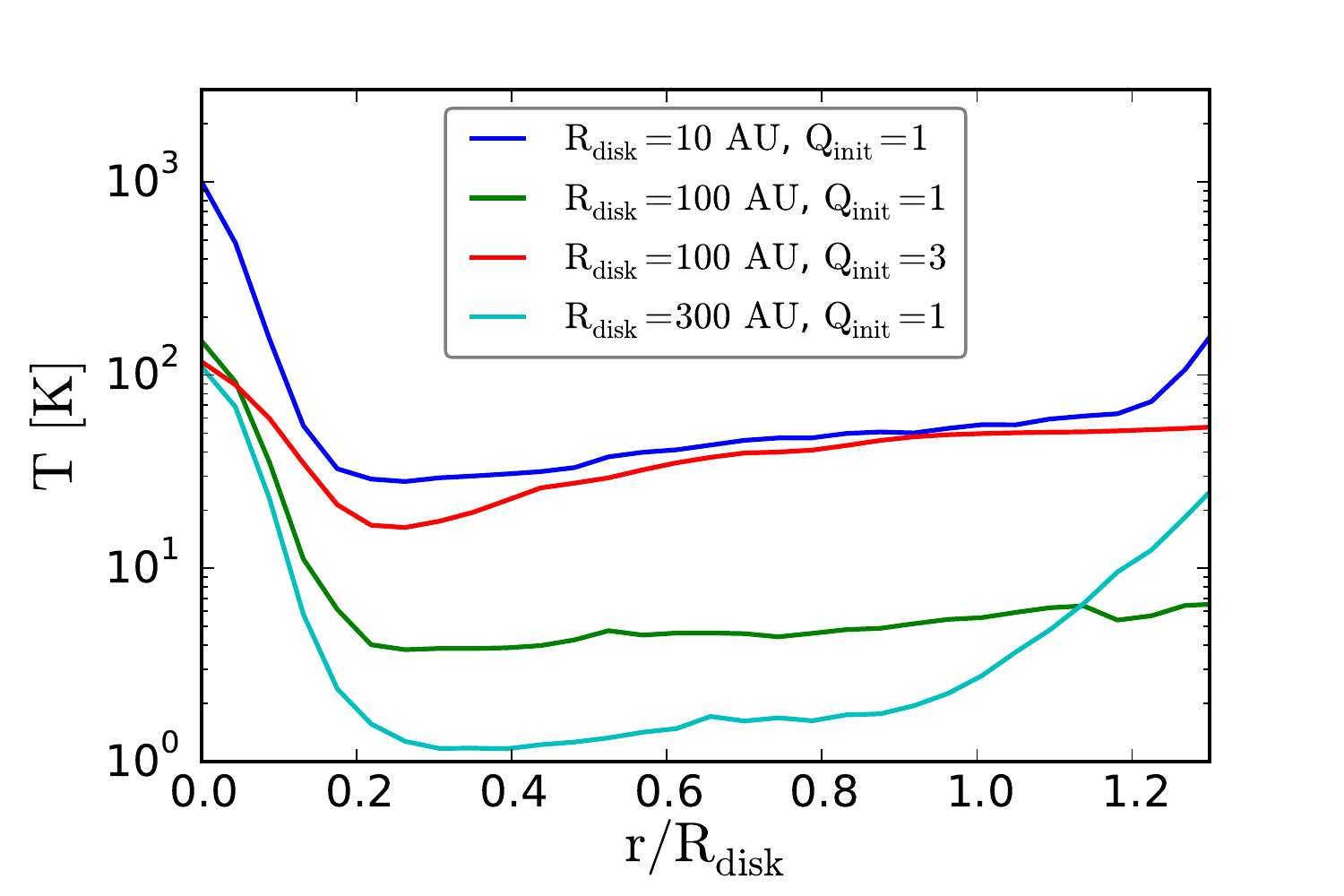}
    \end{subfigure} \\
    \begin{subfigure}[b]{0.5\textwidth}
        \includegraphics[width=\textwidth]{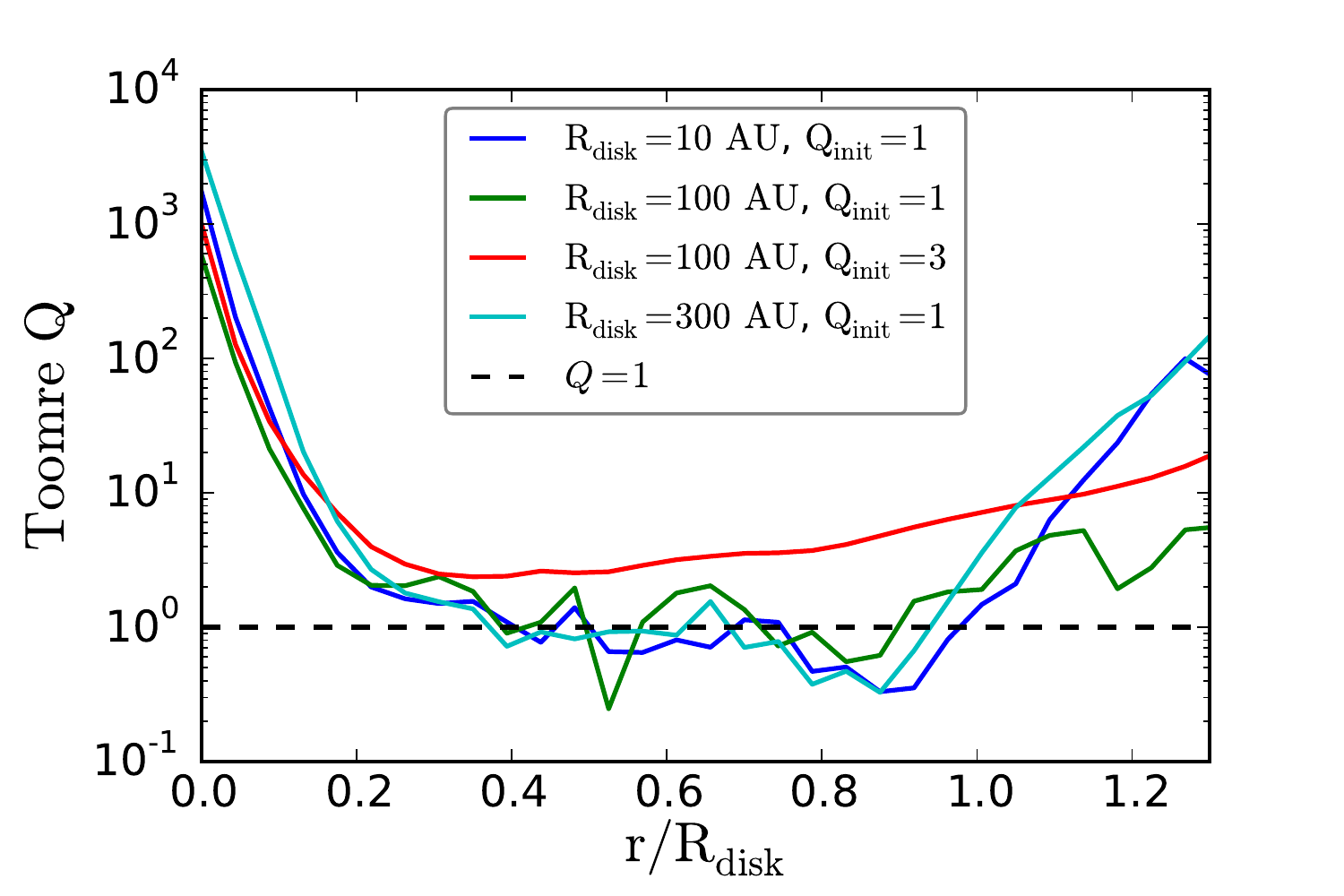}
    \end{subfigure} 
    \caption{Radial profiles at $t = 1.5 \, T_{disk}$ for temperature $T$ and effective Toomre $Q$ parameter for all simulations with the highest resolution on the finest grid, normalized in radial direction. The simulation with $Q_{init}=3$ is added for comparison. Again, we see a different behavior of the simulation with $R_{disk}=100$ AU and $Q_{init}=1$, displaying a smaller increase in temperature because of its smoother transition zone. The dips where the $Q$ parameter goes below 1 are consistent with the density peaks in figures \ref{fig:PS_DensityVelocity} and \ref{fig:Evolved_Structures}.}\label{fig:PS_TemperatureToomreQ}
\end{figure}

In this section we study the influence of the disk extension on the resolution dependent outcome of our model. In general, observations of protoplanetary disks show a wide variety of disk morphologies \citep[e.g.,][]{2014ApJ...781...87A,2014A&A...568A..40G}. Therefore, disk characteristics are likely to vary in radial extent. Additionally, some exotic systems like NN Ser are expected to host very compact systems of only a few AU radius \citep{2014A&A...563A..61S}. Thus, we test our setup on its ability to model disks of varying radius. Note that all our simulated disks have the same overall mass $M_{disk}$ and because of the $Q$ dependency of our model, the temperature is adjusted to be able to reproduce global $Q$ unstable configurations.

\begin{figure*}[bt]
        \centering
        \begin{subfigure}[b]{0.33\textwidth}
                \includegraphics[width=\textwidth]{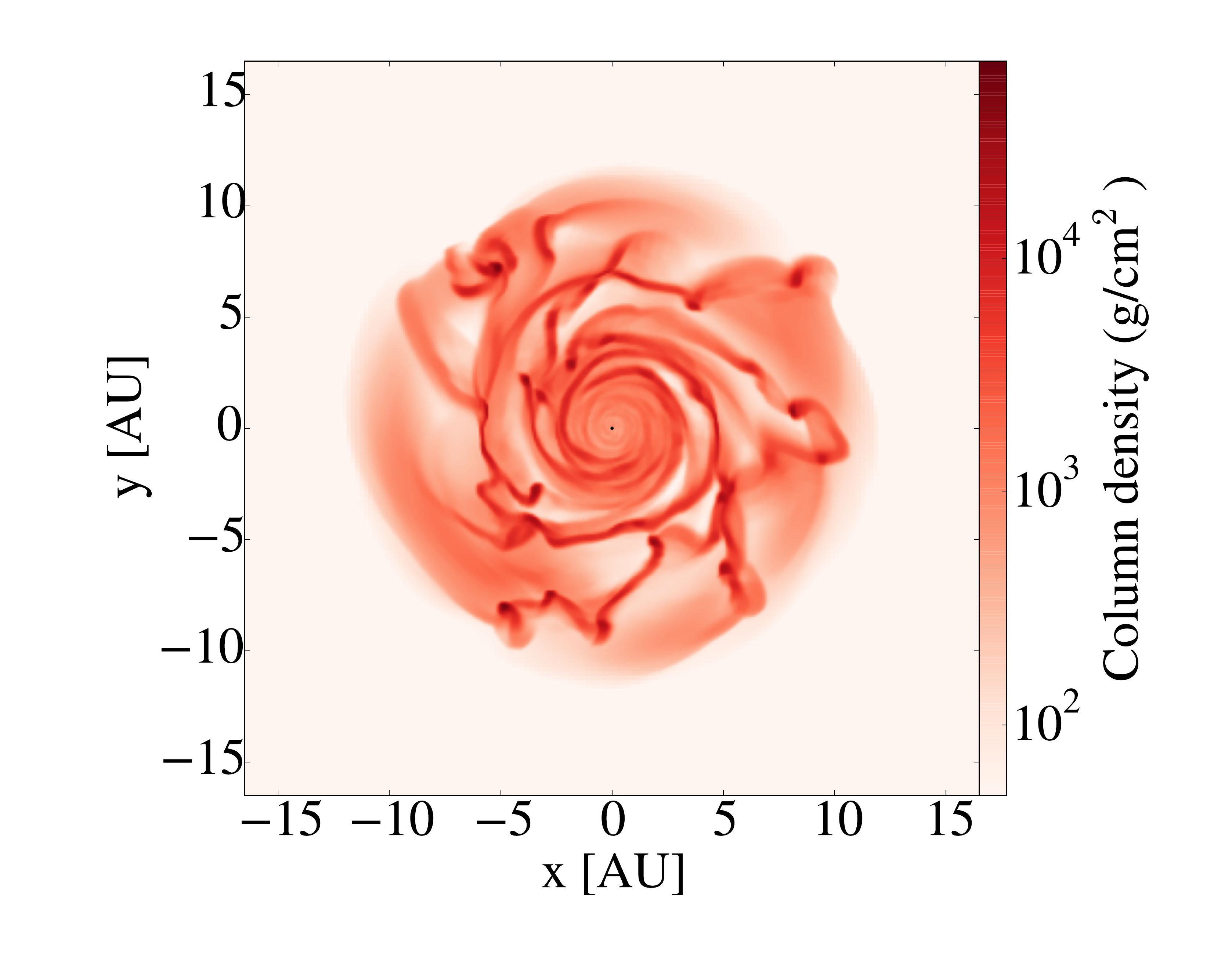}
        \end{subfigure}
        \begin{subfigure}[b]{0.33\textwidth}
                \includegraphics[width=\textwidth]{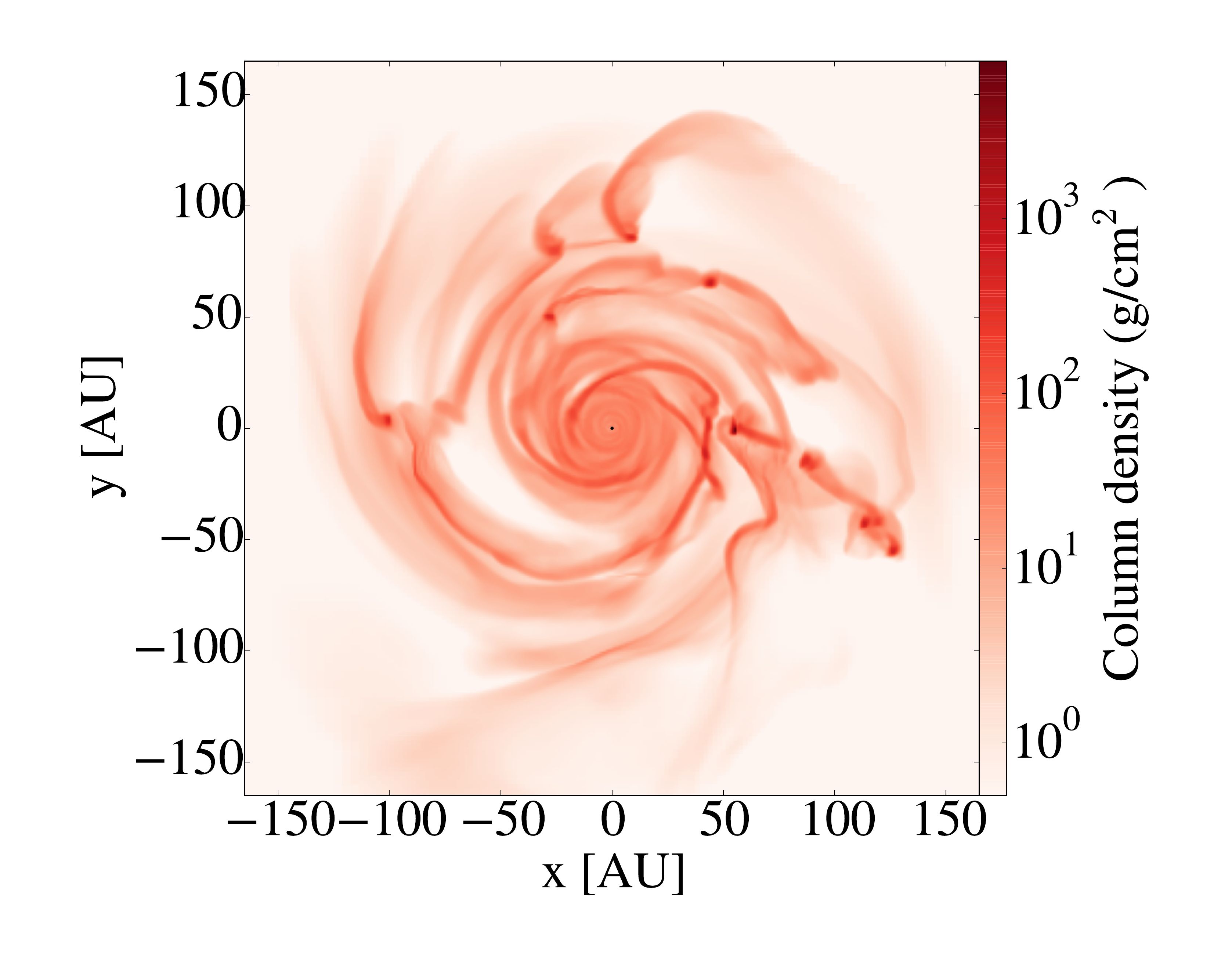}
        \end{subfigure} 
        \begin{subfigure}[b]{0.33\textwidth}
                \includegraphics[width=\textwidth]{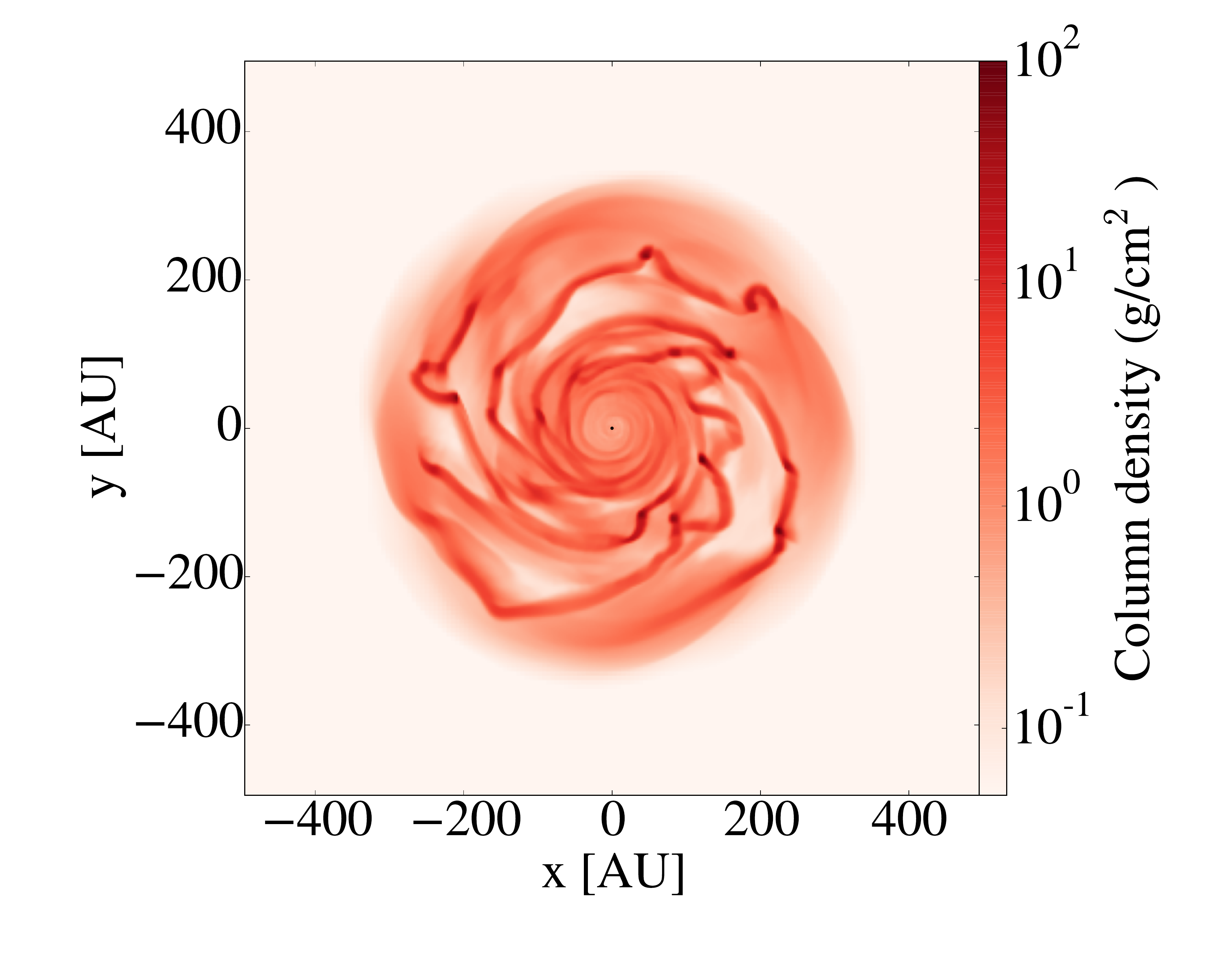}
        \end{subfigure} \\
        \begin{subfigure}[b]{0.33\textwidth}
                \includegraphics[width=\textwidth]{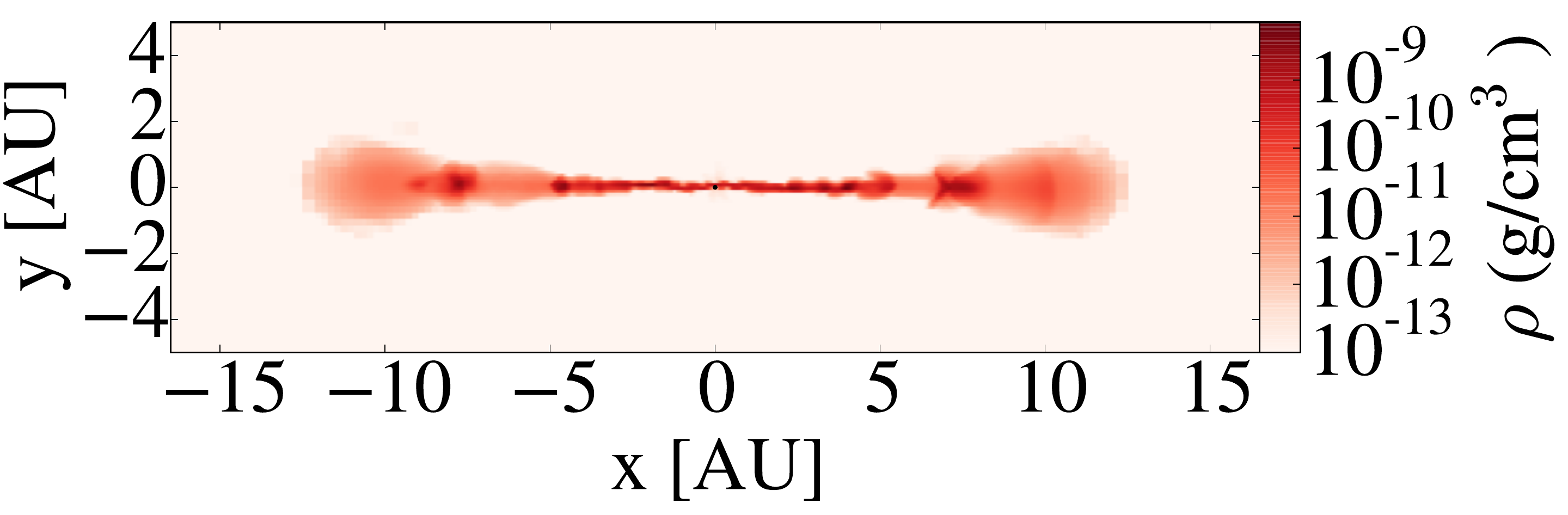}
        \end{subfigure}
        \begin{subfigure}[b]{0.33\textwidth}
                \includegraphics[width=\textwidth]{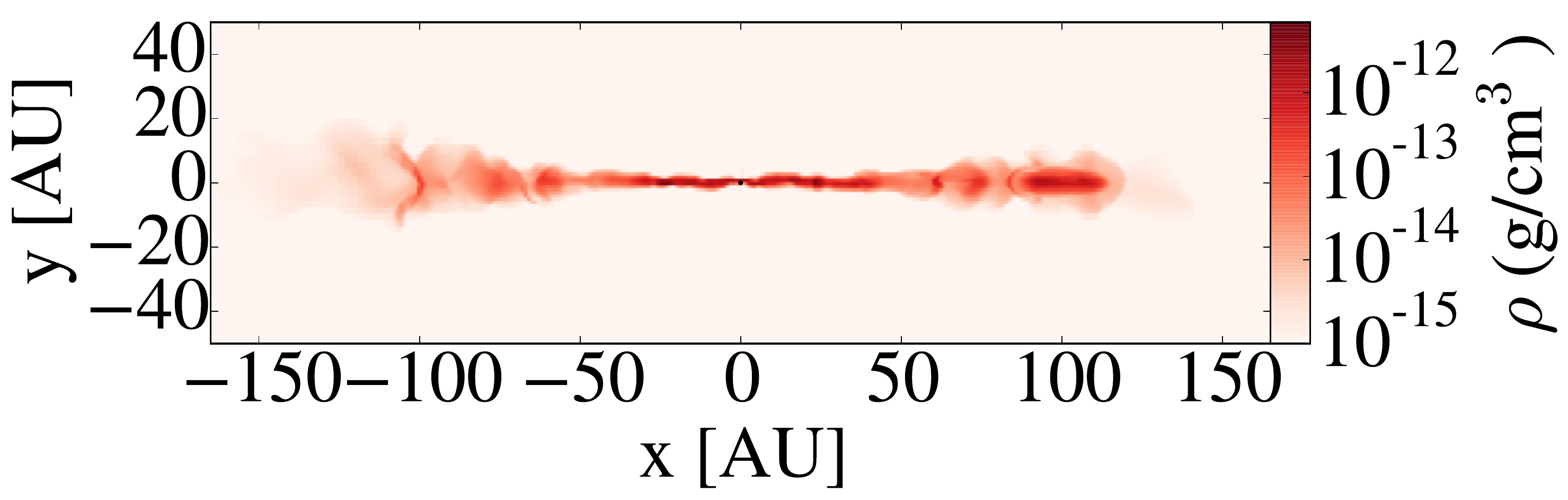}
        \end{subfigure} 
        \begin{subfigure}[b]{0.33\textwidth}
                \includegraphics[width=\textwidth]{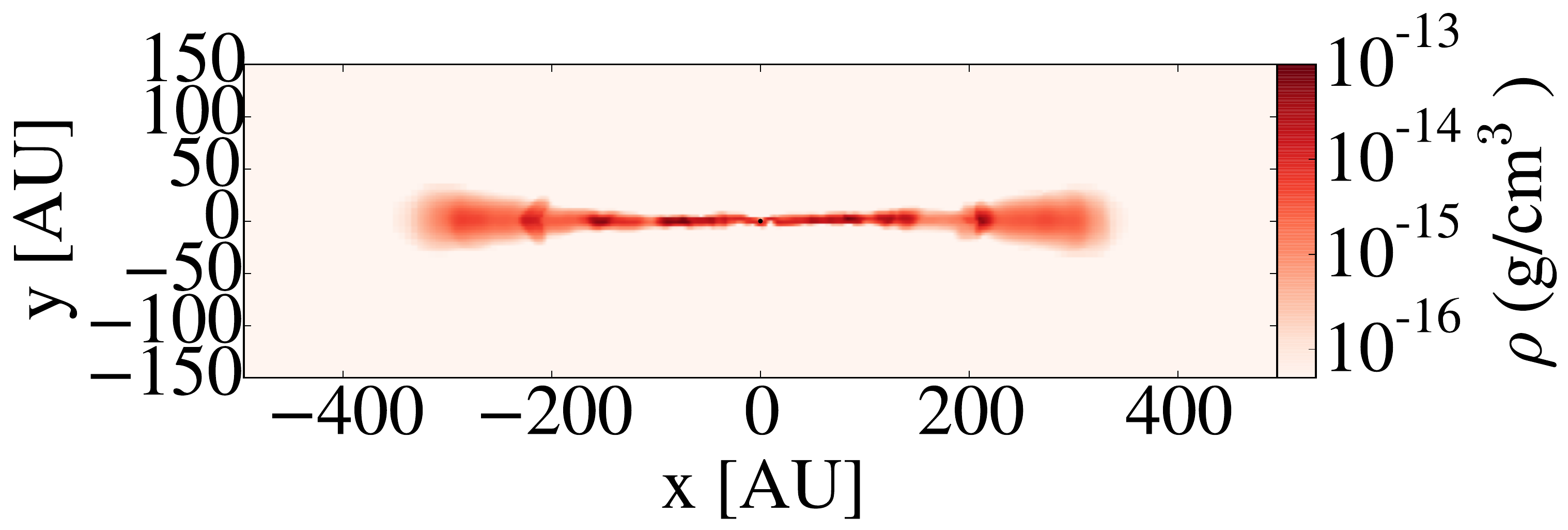}
        \end{subfigure}
        \caption{\textit{Top row}: Density projections of all disks with $Q_{init}=1$ and $l_{max} = 4$, $t = 1.5 \, T_{disk}$. \textit{Bottom row}: Vertical density slices of all disks with $Q_{init}=1$, $l_{max} = 4$ and $t = 1.5 \, T_{disk}$.}\label{fig:Evolved_Structures}
\end{figure*}

Figure \ref{fig:PS_DensityVelocity} and \ref{fig:PS_TemperatureToomreQ} show radial profiles for the parameters $\Sigma$, $v_{rot}$, $T$ and effective $Q$ for all simulations with the highest resolution, i.e., with maximum refinement level $l_{max}=4$. In principle, the simulations with different initial disk radii reproduce similar results and the distributions of the various physical parameters show similar trends. However, there are some differences, which mainly arise because of the single peak in the density distribution of the simulation with $R_{disk}=100$ AU and $Q_{init}=1$ at around 52 AU separation from the central object. This peak, resulting from a vary massive clump (compare Figure \ref{fig:Evolved_Structures}), influences the whole disk structure, and drains material from the other parts of the disk, which results in a different average distribution of matter in the disk, compared to the simulations with $R_{disk}=10$ AU and $R_{disk}=300$ AU. While we expect this peak essentially to represent the stochastic nature of fragmentation, its presence leads to differences for many other parameters as well. For example, the velocity decrease is shallower, especially in comparison with the $R_{disk} = 10$ AU run, which shows a sharp transition at its outer edge from disk material to surrounding material. Addtionally, the temperature distribution shows an approximately constant distribution up to $1.3 \, R_{disk}$, because the (cold) disk material (in comparison with the surrounding gas) is extended over a larger radius and therefore the transition from disk gas to surrounding gas happens at larger radii.

The Toomre $Q$ radial profiles can be matched very well with the profiles for density and temperature. Thus, the most pronounced dip in the $Q$ radial profile is found at approximately $0.5 \, R_{disk}$ separation for the simulation with $R_{disk} = 100$ AU. The simulations with $R_{disk}=10$ AU and $R_{disk}=300$ AU show a very similar (radially) normalized behavior. Their most pronounced $Q$ dips are at separations $0.5 \, R_{disk}$ and $0.8-0.9 \, R_{disk}$. Matching this observation with Figure \ref{fig:Evolved_Structures}, we see that the highest amount of fragmentation can be spotted in these regions.

Figure \ref{fig:Evolved_Structures} shows a final impression of the evolved structures for all simulations. The turbulent structures of the disk can be seen in the column density, as well as in the volume density plots. The projected face-on views provide detailed insight into the fragmented disk structures. The edge-on views represent very well the flared disk structures and reveal the regions of the highest density, where the most massive clumps are located.

\section{Conclusions}
\label{sec:conclusions}

We performed simulations of self-gravitating, massive protoplanetary disks using the AMR code \textsc{enzo}. Our physical setup was motivated by the semianalytical approach of \citet{2014A&A...563A..61S} for modeling the characteristics of a formation via GI in diverse planet formation environments.
We calculated the time evolution of disk configurations close to the $Q_{init}=1$ threshold with disk radii of $R_{disk}=10, 100$ and $300$ AU, varying grid settings, and the time evolution of a stable disk ($Q_{init} = 3$) with $R_{disk}=100$ AU to validate our approach.

The $Q_{init}=1$ disks display the onset of large-scale GI on the orbital timescale, yielding gas fragmentation and the formation of clumps throughout the disk material. The imprint of AMR effects plays a major role in correctly modeling the fragmentation in the disk. Only the highest resolution levels yield pronounced signals of fragmentation and single peaks in the developing spiral arms. Additionally, the structure of the spiral arms and the fragmentation process is qualitatively different for different refinement levels. Whereas for low resolution the spiral arms are less pronounced and the forming clumps differ only slightly from their surrounding medium, with higher resolution the clumps build sharp column density peaks. 

To summarize these findings, we are able to define a lower limit for the resolution in global AMR simulations of GI induced fragmentation in self-gravitating protoplanetary disks. A comparison of the imprint of the maximum refinement level on the fragmentation structure and the building of clumps (figures \ref{fig:Q3_prj} and \ref{fig:Clumps}) yields a minimum ratio of fragmenting disk radius $R_{disk}$ to resolution level $r_{lvl}$ (in physical units). Thus, to induce the development of spiral arms in the disk this ratio has to be
\begin{align}
R_{disk} / r_{lvl} \gtrapprox 50,
\end{align}
corresponding to our simulations with $l_{max} = 3$.
However, to be able to reveal further fragmentation and the building of distinct clumps in the disk spiral structures and resolve the planetary wake around them, one has to satisfy the more rigid criterion of
\begin{align}
R_{disk} / r_{lvl} \gtrapprox 100,
\end{align}
which corresponds to our simulations with $l_{max} = 4$. These critera, giving the number of horizontal cells the disk radius should contain, should be understood as a lower limit to resolve fragmentation structures in a global disk view. Running higher resolutions will most likely yield even better and more resolved structures.

As another valuable feature of the 3D implementation of our setup, disks of all radii conserve their flared structure very well and resolve deviations in scale height during the fragmentation process. This might be especially important to investigate deviations from vertical and axisymmetry in the planet formation process. This is in agreement with the findings of \citet{2008ASPC..398..243M}, who underline the importance of (initial) resolution in the vertical direction to resolve clumping effects. Even if it is in general hard to compare the outcome of SPH calculations with those from grid-based models, our resolution dependent results are similar to those of \citet{2006MNRAS.373.1039N} concerning the importance of vertical resolution in SPH simulations.

The total angular momentum in all simulations is approximately conserved (with total errors of up to 2\%) during the time evolution of the simulations. The marginal loss in angular momentum, shown in Figure \ref{fig:TotalAM}, can be explained with discretization effects due to the grid structure of the simulations. From a numerical point of view the simulations presented here can be understood as a first step toward a more realistic coverage of turbulence effects in accretion disk scenarios with high Mach numbers. As found by \citet{2011ApJ...731...62F} the energy injection scale of gravity-driven turbulence is close to the local Jeans scale. Therefore, in AMR simulations in which gravitational energy is converted into turbulent energy, it is crucial to resolve the Jeans length by (at least) 32 \citep{2011ApJ...731...62F} or even 64 \citep{2013MNRAS.430..588L,2013MNRAS.432..668L} cells. Therefore, the maximum refinement level chosen in our simulations is a constraint for even better angular momentum conservation.%

Additional ideas to speed up simulations of self-gravitating accretion disks and to enable simulations on a wider range of spatial scales are necessary to eventually cover the range from the global disk view down to smaller scales as, e.g., the physics of circumplanetary disks \citep{2013DPS....4551005M}. A first step toward this goal might be the introduction of a more specific refinement criterion, which only covers areas that can be associated with long-lasting clumps, which might eventually end up on a planetary mass scale. 

A particular interesting feature of the simulations from a physical point of view is the formation and rapid inward migration of clumps. They migrate inward within a few orbital times, which is roughly consistent with the timescale of Type I migration \citep{2014prpl.conf..643H}. Apart from the relatively small masses, the rapid inward migration is consistent with the findings of \citet{2011ApJ...737L..42M} and \cite{2011MNRAS.416.1971B}. The migration in unstable and turbulent disks is driven by several forces, such as clump-disk interactions (like Lindblad torques) or stochastic torques. Identifying the major driving forces in our scenario and to find a stabilization mechanism for the migrating clumps are subjects for further research. 

To overcome these fast migration scenarios, \citet{2014arXiv1411.5366M} proposed the sculpting of long-lasting dust rings through pressure maxima in the disk. However, a scenario to explain the migration stalling purely based on the thermal evolution of the gas might be the influence of radiative feedback from the central star. Following \cite{1997ApJ...490..368C}, the photoheating would transform a significant part of the disk to be stabilized against fragmentation instabilities \citep{2014A&A...563A..61S}. Clumps migrating toward such stabilized regions might still be able to grow and evacuate the disk in this region. This stalling of the inward migration and possible opening of gaps in the disk would prevent them from being destroyed by tidal interactions \citep{2012ApJ...746..110Z}.

The timescale of the disk's evolution, comparable to the orbital timescale, and the fast fragmentation and formation of clumps is of major interest for the formation of planets in exotic systems like the NN Serpentis binary system. Since its evolutionary timescale is determined by the cooling age of its white dwarf \citep{2011AIPC.1331..281H}, the formation of the planets must happen on a timescale of $\lessapprox 10^6$ yr. Thus, the rapid fragmentation and clumping opens up a possible way to create the proposed NN Ser planets in the appropriate time. However, further observational coverage and more realistic theoretical investigations of this system are needed.

We have presented here a systematic study exploring the GI in self-gravitating protoplanetary disks with AMR simulations. We expect that this technique can provide additional insight into the formation of massive self-gravitating clumps in future simulations, which will complement the existing numerical approaches well. This technique can be combined with additional physics modules like the chemistry package \textsc{krome} \citep{2014MNRAS.439.2386G}, radiation transport techniques \citep{2011MNRAS.414.3458W} and a sink particle algorithm \citep{2013MNRAS.436.2989L} for an improved modeling of the formation of planets. The simulations presented here will be particularly valuable as a reference model to understand the evolution of the GI in the absence of additional processes that can complicate the situation, but already shed light on the important role and necessity of high resolution in grid-based calculations.

\begin{acknowledgements}
We thank the anonymous referee for valuable suggestions, which helped to improve the quality of the paper. Moreover, we thank Stefan Dreizler, Robi Banerjee, Rick Hessman, Farzana Meru, Marcel V\"olschow, Christiane Diehl and Muhammad Latif for stimulating discussions. The computer simulations featured in this work were run on the {\em GWDG} computing cluster (\textsc{gwdg.de}). The simulation results were analyzed using the visualization toolkit for astrophysical data \textsc{yt} \citep[][\textsc{yt-project.org}]{2011ApJS..192....9T}, which is based on the \textsc{matplotlib} Python library \citep[][\textsc{matplotlib.org}]{Hunter:2007}.
\end{acknowledgements}


\begin{thebibliography}{66}
\expandafter\ifx\csname natexlab\endcsname\relax\def\natexlab#1{#1}\fi

\bibitem[{{Agertz} {et~al.}(2007){Agertz}, {Moore}, {Stadel}, {Potter},
  {Miniati}, {Read}, {Mayer}, {Gawryszczak}, {Kravtsov}, {Nordlund}, {Pearce},
  {Quilis}, {Rudd}, {Springel}, {Stone}, {Tasker}, {Teyssier}, {Wadsley}, \&
  {Walder}}]{2007MNRAS.380..963A}
{Agertz}, O., {Moore}, B., {Stadel}, J., {et~al.} 2007, \mnras, 380, 963

\bibitem[{{Avenhaus} {et~al.}(2014){Avenhaus}, {Quanz}, {Schmid}, {Meyer},
  {Garufi}, {Wolf}, \& {Dominik}}]{2014ApJ...781...87A}
{Avenhaus}, H., {Quanz}, S.~P., {Schmid}, H.~M., {et~al.} 2014, \apj, 781, 87

\bibitem[{{Baruteau} {et~al.}(2011){Baruteau}, {Meru}, \&
  {Paardekooper}}]{2011MNRAS.416.1971B}
{Baruteau}, C., {Meru}, F., \& {Paardekooper}, S.-J. 2011, \mnras, 416, 1971

\bibitem[{{Benz} {et~al.}(2014){Benz}, {Ida}, {Alibert}, {Lin}, \&
  {Mordasini}}]{2014prpl.conf..691B}
{Benz}, W., {Ida}, S., {Alibert}, Y., {Lin}, D., \& {Mordasini}, C. 2014,
  Protostars and Planets VI, 691

\bibitem[{{Berger} \& {Colella}(1989)}]{1989JCoPh..82...64B}
{Berger}, M.~J. \& {Colella}, P. 1989, Journal of Computational Physics, 82, 64

\bibitem[{{Beuermann} {et~al.}(2013){Beuermann}, {Dreizler}, \&
  {Hessman}}]{2013A&A...555A.133B}
{Beuermann}, K., {Dreizler}, S., \& {Hessman}, F.~V. 2013, \aap, 555, A133

\bibitem[{{Bodenheimer} {et~al.}(2000){Bodenheimer}, {Hubickyj}, \&
  {Lissauer}}]{2000Icar..143....2B}
{Bodenheimer}, P., {Hubickyj}, O., \& {Lissauer}, J.~J. 2000, \icarus, 143, 2

\bibitem[{{Boley}(2009)}]{2009ApJ...695L..53B}
{Boley}, A.~C. 2009, \apjl, 695, L53

\bibitem[{{Boley} {et~al.}(2007){Boley}, {Durisen}, {Nordlund}, \&
  {Lord}}]{2007ApJ...665.1254B}
{Boley}, A.~C., {Durisen}, R.~H., {Nordlund}, {\AA}., \& {Lord}, J. 2007, \apj,
  665, 1254

\bibitem[{{Borucki} {et~al.}(2010){Borucki}, {Koch}, {Basri}, {Batalha},
  {Brown}, {Caldwell}, {Caldwell}, {Christensen-Dalsgaard}, {Cochran},
  {DeVore}, {Dunham}, {Dupree}, {Gautier}, {Geary}, {Gilliland}, {Gould},
  {Howell}, {Jenkins}, {Kondo}, {Latham}, {Marcy}, {Meibom}, {Kjeldsen},
  {Lissauer}, {Monet}, {Morrison}, {Sasselov}, {Tarter}, {Boss}, {Brownlee},
  {Owen}, {Buzasi}, {Charbonneau}, {Doyle}, {Fortney}, {Ford}, {Holman},
  {Seager}, {Steffen}, {Welsh}, {Rowe}, {Anderson}, {Buchhave}, {Ciardi},
  {Walkowicz}, {Sherry}, {Horch}, {Isaacson}, {Everett}, {Fischer}, {Torres},
  {Johnson}, {Endl}, {MacQueen}, {Bryson}, {Dotson}, {Haas}, {Kolodziejczak},
  {Van Cleve}, {Chandrasekaran}, {Twicken}, {Quintana}, {Clarke}, {Allen},
  {Li}, {Wu}, {Tenenbaum}, {Verner}, {Bruhweiler}, {Barnes}, \&
  {Prsa}}]{2010Sci...327..977B}
{Borucki}, W.~J., {Koch}, D., {Basri}, G., {et~al.} 2010, Science, 327, 977

\bibitem[{{Boss}(1997)}]{1997Sci...276.1836B}
{Boss}, A.~P. 1997, Science, 276, 1836

\bibitem[{{Boss}(2003)}]{2003ApJ...599..577B}
{Boss}, A.~P. 2003, \apj, 599, 577

\bibitem[{{Bryan} {et~al.}(2014){Bryan}, {Norman}, {O'Shea}, {Abel}, {Wise},
  {Turk}, {Reynolds}, {Collins}, {Wang}, {Skillman}, {Smith}, {Harkness},
  {Bordner}, {Kim}, {Kuhlen}, {Xu}, {Goldbaum}, {Hummels}, {Kritsuk}, {Tasker},
  {Skory}, {Simpson}, {Hahn}, {Oishi}, {So}, {Zhao}, {Cen}, {Li}, \& {Enzo
  Collaboration}}]{2014ApJS..211...19B}
{Bryan}, G.~L., {Norman}, M.~L., {O'Shea}, B.~W., {et~al.} 2014, \apjs, 211, 19

\bibitem[{{Bryan} {et~al.}(1995){Bryan}, {Norman}, {Stone}, {Cen}, \&
  {Ostriker}}]{1995CoPhC..89..149B}
{Bryan}, G.~L., {Norman}, M.~L., {Stone}, J.~M., {Cen}, R., \& {Ostriker},
  J.~P. 1995, Computer Physics Communications, 89, 149

\bibitem[{{Chiang} \& {Goldreich}(1997)}]{1997ApJ...490..368C}
{Chiang}, E.~I. \& {Goldreich}, P. 1997, \apj, 490, 368

\bibitem[{{Colella} \& {Woodward}(1984)}]{1984JCoPh..54..174C}
{Colella}, P. \& {Woodward}, P.~R. 1984, Journal of Computational Physics, 54,
  174

\bibitem[{{Federrath} {et~al.}(2011){Federrath}, {Sur}, {Schleicher},
  {Banerjee}, \& {Klessen}}]{2011ApJ...731...62F}
{Federrath}, C., {Sur}, S., {Schleicher}, D.~R.~G., {Banerjee}, R., \&
  {Klessen}, R.~S. 2011, \apj, 731, 62

\bibitem[{{Garufi} {et~al.}(2014){Garufi}, {Quanz}, {Schmid}, {Avenhaus},
  {Buenzli}, \& {Wolf}}]{2014A&A...568A..40G}
{Garufi}, A., {Quanz}, S.~P., {Schmid}, H.~M., {et~al.} 2014, \aap, 568, A40

\bibitem[{{Gawryszczak} {et~al.}(2010){Gawryszczak}, {Mayer}, {Boley}, \&
  {Tasker}}]{2010EAS....42..267G}
{Gawryszczak}, A.~J., {Mayer}, L., {Boley}, A., \& {Tasker}, E. 2010, in EAS
  Publications Series, Vol.~42, EAS Publications Series, ed.
  K.~{Go{\.z}dziewski}, A.~{Niedzielski}, \& J.~{Schneider}, 267--274

\bibitem[{{Grassi} {et~al.}(2014){Grassi}, {Bovino}, {Schleicher}, {Prieto},
  {Seifried}, {Simoncini}, \& {Gianturco}}]{2014MNRAS.439.2386G}
{Grassi}, T., {Bovino}, S., {Schleicher}, D.~R.~G., {et~al.} 2014, \mnras, 439,
  2386

\bibitem[{{Haefner} {et~al.}(2004){Haefner}, {Fiedler}, {Butler}, \&
  {Barwig}}]{2004A&A...428..181H}
{Haefner}, R., {Fiedler}, A., {Butler}, K., \& {Barwig}, H. 2004, \aap, 428,
  181

\bibitem[{{Helled} {et~al.}(2014){Helled}, {Bodenheimer}, {Podolak}, {Boley},
  {Meru}, {Nayakshin}, {Fortney}, {Mayer}, {Alibert}, \&
  {Boss}}]{2014prpl.conf..643H}
{Helled}, R., {Bodenheimer}, P., {Podolak}, M., {et~al.} 2014, Protostars and
  Planets VI, 643

\bibitem[{{Hessman} {et~al.}(2011){Hessman}, {Beuermann}, {Dreizler}, {Marsh},
  {Parsons}, {Copperwheat}, {Winget}, {Miller}, {Hermes}, {Schreiber}, {Kley},
  {Dhillon}, \& {Littlefair}}]{2011AIPC.1331..281H}
{Hessman}, F.~V., {Beuermann}, K., {Dreizler}, S., {et~al.} 2011, in American
  Institute of Physics Conference Series, Vol. 1331, American Institute of
  Physics Conference Series, ed. S.~{Schuh}, H.~{Drechsel}, \& U.~{Heber},
  281--286

\bibitem[{{Hockney} \& {Eastwood}(1988)}]{1988csup.book.....H}
{Hockney}, R.~W. \& {Eastwood}, J.~W. 1988, {Computer simulation using
  particles}

\bibitem[{{Hopkins}(2013)}]{2013MNRAS.428.2840H}
{Hopkins}, P.~F. 2013, \mnras, 428, 2840

\bibitem[{{Horner} {et~al.}(2012){Horner}, {Wittenmyer}, {Hinse}, \&
  {Tinney}}]{2012MNRAS.425..749H}
{Horner}, J., {Wittenmyer}, R.~A., {Hinse}, T.~C., \& {Tinney}, C.~G. 2012,
  \mnras, 425, 749

\bibitem[{Hunter(2007)}]{Hunter:2007}
Hunter, J.~D. 2007, Computing In Science \& Engineering, 9, 90

\bibitem[{{Inayoshi} \& {Haiman}(2014)}]{2014MNRAS.445.1549I}
{Inayoshi}, K. \& {Haiman}, Z. 2014, \mnras, 445, 1549

\bibitem[{{Johnson} {et~al.}(2012){Johnson}, {Mousis}, {Lunine}, \&
  {Madhusudhan}}]{2012EGUGA..1411532J}
{Johnson}, T.~V., {Mousis}, O., {Lunine}, J.~I., \& {Madhusudhan}, N. 2012, in
  EGU General Assembly Conference Abstracts, Vol.~14, EGU General Assembly
  Conference Abstracts, ed. A.~{Abbasi} \& N.~{Giesen}, 11532

\bibitem[{{Latif} \& {Schleicher}(2015)}]{2015MNRAS.449...77L}
{Latif}, M.~A. \& {Schleicher}, D.~R.~G. 2015, \mnras, 449, 77

\bibitem[{{Latif} {et~al.}(2013{\natexlab{a}}){Latif}, {Schleicher}, {Schmidt},
  \& {Niemeyer}}]{2013MNRAS.430..588L}
{Latif}, M.~A., {Schleicher}, D.~R.~G., {Schmidt}, W., \& {Niemeyer}, J.
  2013{\natexlab{a}}, \mnras, 430, 588

\bibitem[{{Latif} {et~al.}(2013{\natexlab{b}}){Latif}, {Schleicher}, {Schmidt},
  \& {Niemeyer}}]{2013MNRAS.432..668L}
{Latif}, M.~A., {Schleicher}, D.~R.~G., {Schmidt}, W., \& {Niemeyer}, J.
  2013{\natexlab{b}}, \mnras, 432, 668

\bibitem[{{Latif} {et~al.}(2013{\natexlab{c}}){Latif}, {Schleicher}, {Schmidt},
  \& {Niemeyer}}]{2013MNRAS.436.2989L}
{Latif}, M.~A., {Schleicher}, D.~R.~G., {Schmidt}, W., \& {Niemeyer}, J.~C.
  2013{\natexlab{c}}, \mnras, 436, 2989

\bibitem[{{Lemaster} \& {Stone}(2009)}]{2009ApJ...691.1092L}
{Lemaster}, M.~N. \& {Stone}, J.~M. 2009, \apj, 691, 1092

\bibitem[{{Lodato}(2008)}]{2008NewAR..52...21L}
{Lodato}, G. 2008, \nar, 52, 21

\bibitem[{{Marsh} {et~al.}(2014){Marsh}, {Parsons}, {Bours}, {Littlefair},
  {Copperwheat}, {Dhillon}, {Breedt}, {Caceres}, \&
  {Schreiber}}]{2014MNRAS.437..475M}
{Marsh}, T.~R., {Parsons}, S.~G., {Bours}, M.~C.~P., {et~al.} 2014, \mnras,
  437, 475

\bibitem[{{Mayer} \& {Gawryszczak}(2008)}]{2008ASPC..398..243M}
{Mayer}, L. \& {Gawryszczak}, A.~J. 2008, in Astronomical Society of the
  Pacific Conference Series, Vol. 398, Extreme Solar Systems, ed. D.~{Fischer},
  F.~A. {Rasio}, S.~E. {Thorsett}, \& A.~{Wolszczan}, 243

\bibitem[{{Mayer} {et~al.}(2007){Mayer}, {Lufkin}, {Quinn}, \&
  {Wadsley}}]{2007ApJ...661L..77M}
{Mayer}, L., {Lufkin}, G., {Quinn}, T., \& {Wadsley}, J. 2007, \apjl, 661, L77

\bibitem[{{Mayer} {et~al.}(2003){Mayer}, {Quinn}, {Wadsley}, \&
  {Stadel}}]{2003ASPC..294..281M}
{Mayer}, L., {Quinn}, T., {Wadsley}, J., \& {Stadel}, J. 2003, in Astronomical
  Society of the Pacific Conference Series, Vol. 294, Scientific Frontiers in
  Research on Extrasolar Planets, ed. D.~{Deming} \& S.~{Seager}, 281--286

\bibitem[{{Meru} \& {Bate}(2011)}]{2011MNRAS.410..559M}
{Meru}, F. \& {Bate}, M.~R. 2011, \mnras, 410, 559

\bibitem[{{Meru} {et~al.}(2014){Meru}, {Quanz}, {Reggiani}, {Baruteau}, \&
  {Pineda}}]{2014arXiv1411.5366M}
{Meru}, F., {Quanz}, S.~P., {Reggiani}, M., {Baruteau}, C., \& {Pineda}, J.~E.
  2014, ArXiv e-prints [\eprint[arXiv]{1411.5366}]

\bibitem[{{Mestel}(1963)}]{1963MNRAS.126..553M}
{Mestel}, L. 1963, \mnras, 126, 553

\bibitem[{{Michael} {et~al.}(2011){Michael}, {Durisen}, \&
  {Boley}}]{2011ApJ...737L..42M}
{Michael}, S., {Durisen}, R.~H., \& {Boley}, A.~C. 2011, \apjl, 737, L42

\bibitem[{{Morbidelli} {et~al.}(2012){Morbidelli}, {Lunine}, {O'Brien},
  {Raymond}, \& {Walsh}}]{2012AREPS..40..251M}
{Morbidelli}, A., {Lunine}, J.~I., {O'Brien}, D.~P., {Raymond}, S.~N., \&
  {Walsh}, K.~J. 2012, Annual Review of Earth and Planetary Sciences, 40, 251

\bibitem[{{Morbidelli} {et~al.}(2013){Morbidelli}, {Szulagyi}, {Crida},
  {Tanigawa}, {Lega}, {Masset}, \& {Bitsch}}]{2013DPS....4551005M}
{Morbidelli}, A., {Szulagyi}, J., {Crida}, A., {et~al.} 2013, in AAS/Division
  for Planetary Sciences Meeting Abstracts, Vol.~45, AAS/Division for Planetary
  Sciences Meeting Abstracts, 510.05

\bibitem[{{Nelson}(2006)}]{2006MNRAS.373.1039N}
{Nelson}, A.~F. 2006, \mnras, 373, 1039

\bibitem[{{O'Shea} {et~al.}(2004){O'Shea}, {Bryan}, {Bordner}, {Norman},
  {Abel}, {Harkness}, \& {Kritsuk}}]{2004astro.ph..3044O}
{O'Shea}, B.~W., {Bryan}, G., {Bordner}, J., {et~al.} 2004, ArXiv Astrophysics
  e-prints [\eprint{astro-ph/0403044}]

\bibitem[{{Paardekooper} \& {Mellema}(2006)}]{2006A&A...450.1203P}
{Paardekooper}, S.-J. \& {Mellema}, G. 2006, \aap, 450, 1203

\bibitem[{{Parsons} {et~al.}(2014){Parsons}, {Marsh}, {Bours}, {Littlefair},
  {Copperwheat}, {Dhillon}, {Breedt}, {Caceres}, \&
  {Schreiber}}]{2014MNRAS.438L..91P}
{Parsons}, S.~G., {Marsh}, T.~R., {Bours}, M.~C.~P., {et~al.} 2014, \mnras,
  438, L91

\bibitem[{{Price}(2008)}]{2008JCoPh.22710040P}
{Price}, D.~J. 2008, Journal of Computational Physics, 227, 10040

\bibitem[{{Price}(2012)}]{2012JCoPh.231..759P}
{Price}, D.~J. 2012, Journal of Computational Physics, 231, 759

\bibitem[{{Price} \& {Federrath}(2010)}]{2010MNRAS.406.1659P}
{Price}, D.~J. \& {Federrath}, C. 2010, \mnras, 406, 1659

\bibitem[{{Raymond} {et~al.}(2006){Raymond}, {Barnes}, \&
  {Kaib}}]{2006ApJ...644.1223R}
{Raymond}, S.~N., {Barnes}, R., \& {Kaib}, N.~A. 2006, \apj, 644, 1223

\bibitem[{{Raymond} {et~al.}(2014){Raymond}, {Kokubo}, {Morbidelli},
  {Morishima}, \& {Walsh}}]{2014prpl.conf..595R}
{Raymond}, S.~N., {Kokubo}, E., {Morbidelli}, A., {Morishima}, R., \& {Walsh},
  K.~J. 2014, Protostars and Planets VI, 595

\bibitem[{{Read} \& {Hayfield}(2012)}]{2012MNRAS.422.3037R}
{Read}, J.~I. \& {Hayfield}, T. 2012, \mnras, 422, 3037

\bibitem[{{Schleicher} {et~al.}(2015){Schleicher}, {Bovino}, {Latif},
  {Ferrara}, \& {Grassi}}]{2015arXiv150406296S}
{Schleicher}, D.~R.~G., {Bovino}, S., {Latif}, M.~A., {Ferrara}, A., \&
  {Grassi}, T. 2015, ArXiv e-prints [\eprint[arXiv]{1504.06296}]

\bibitem[{{Schleicher} \& {Dreizler}(2014)}]{2014A&A...563A..61S}
{Schleicher}, D.~R.~G. \& {Dreizler}, S. 2014, \aap, 563, A61

\bibitem[{{Setiawan} {et~al.}(2010){Setiawan}, {Klement}, {Henning}, {Rix},
  {Rochau}, {Rodmann}, \& {Schulze-Hartung}}]{2010Sci...330.1642S}
{Setiawan}, J., {Klement}, R.~J., {Henning}, T., {et~al.} 2010, Science, 330,
  1642

\bibitem[{{Smith} {et~al.}(2010){Smith}, {Silvia}, {Turk}, {Sigurdsson},
  {O'Shea}, {Shull}, \& {Norman}}]{2010AIPC.1294..110S}
{Smith}, B.~D., {Silvia}, D.~W., {Turk}, M.~J., {et~al.} 2010, in American
  Institute of Physics Conference Series, Vol. 1294, American Institute of
  Physics Conference Series, ed. D.~J. {Whalen}, V.~{Bromm}, \& N.~{Yoshida},
  110--115

\bibitem[{{Toomre}(1964)}]{1964ApJ...139.1217T}
{Toomre}, A. 1964, \apj, 139, 1217

\bibitem[{{Truelove} {et~al.}(1998){Truelove}, {Klein}, {McKee}, {Holliman},
  {Howell}, {Greenough}, \& {Woods}}]{1998ApJ...495..821T}
{Truelove}, J.~K., {Klein}, R.~I., {McKee}, C.~F., {et~al.} 1998, \apj, 495,
  821

\bibitem[{{Turk} {et~al.}(2011){Turk}, {Smith}, {Oishi}, {Skory}, {Skillman},
  {Abel}, \& {Norman}}]{2011ApJS..192....9T}
{Turk}, M.~J., {Smith}, B.~D., {Oishi}, J.~S., {et~al.} 2011, \apjs, 192, 9

\bibitem[{{V{\"o}lschow} {et~al.}(2014){V{\"o}lschow}, {Banerjee}, \&
  {Hessman}}]{2014A&A...562A..19V}
{V{\"o}lschow}, M., {Banerjee}, R., \& {Hessman}, F.~V. 2014, \aap, 562, A19

\bibitem[{{Wise} \& {Abel}(2011)}]{2011MNRAS.414.3458W}
{Wise}, J.~H. \& {Abel}, T. 2011, \mnras, 414, 3458

\bibitem[{{Zhu} {et~al.}(2012){Zhu}, {Hartmann}, {Nelson}, \&
  {Gammie}}]{2012ApJ...746..110Z}
{Zhu}, Z., {Hartmann}, L., {Nelson}, R.~P., \& {Gammie}, C.~F. 2012, \apj, 746,
  110

\bibitem[{{Zorotovic} \& {Schreiber}(2013)}]{2013A&A...549A..95Z}
{Zorotovic}, M. \& {Schreiber}, M.~R. 2013, \aap, 549, A95

\end{thebibliography}




\balance

\end{document}